\documentclass{IEEEtran}
\usepackage{ifpdf}
\usepackage{amsmath}
\usepackage{amsfonts}
\usepackage{mathrsfs}
\usepackage{color}
\usepackage{graphicx}
\usepackage{subfigure}
\usepackage{epstopdf}
\usepackage{lineno}
\usepackage{cite}
\usepackage{enumerate}

\newtheorem{theorem}{\textbf{Theorem}}
\newtheorem{lemma}{\textbf{Lemma}}

\newtheorem{corollary}{\textbf{Corollary}}

\newtheorem{definition}{\textbf{Definition}}

\newtheorem{remark}{Remark}
\newtheorem{assumption}{Assumption}

\newcommand{\diag}{\mathrm{diag}}

\graphicspath{{figures/}}

\begin{document}
%
\title{Leader-Driven Opinion Dynamics in Signed Social Networks With Asynchronous Trust/Distrust Level Evolution}
%
%
%

\author{Lei Shi,~\IEEEmembership{Member, IEEE}, Yuhua Cheng,~\IEEEmembership{Senior Member, IEEE}, Jinliang Shao, Xiaofan Wang,~\IEEEmembership{Senior Member, IEEE}, and Hanmin Sheng,~\IEEEmembership{Member, IEEE}
\thanks{This research was supported in part by the National Science Foundation of China under Grant U1830207, Grant 61772003, and Grant 61903066; in part by the Sichuan Science and Technology Program under Grant 2021YFH0042; and in part by the funding from Shenzhen Institute of Artificial Intelligence and Robotics for Society.}
\thanks{Lei Shi, Jinliang Shao and Hanmin Sheng are with the School of Automation Engineering, University of Electronic Science and Technology of China, Chengdu 611731, China, and are also with the Research Center on Crowd Spectrum Intelligence, Shenzhen Institute of Artificial Intelligence and Robotics for Society, Shenzhen 518054, China (email: shilei@uestc.edu.cn (L. Shi), jinliangshao@uestc.edu.cn (J. Shao), shenghanmin@hotmail.com (H. Sheng)).}
\thanks{Yuhua Cheng is with the School of Automation Engineering, University of Electronic Science and Technology of China, Sichuan, 611731, P.~R.~China (email: chengyuhua\_auto@uestc.edu.cn).}
\thanks{Xiaofan Wang is with the School of Mechatronic Engineering and Automation, Shanghai University, Shanghai 200444, P. R. China  (email: xfwang@shu.edu.cn ).}
}

\maketitle

\begin{abstract}
Trust and distrust are common in the opinion interactions among agents in social networks, and they are described by the edges with positive and negative weights in the signed digraph, respectively. It has been shown in social psychology that although the opinions of most agents (followers) tend to prevail, sometimes one agent (leader) with a firm stand and strong influence can impact or even overthrow the preferences of followers. This paper aims to analyze how the leader influences the formation of followers' opinions in signed social networks. In addition, this paper considers an asynchronous evolution mechanism of trust/distrust level based on opinion difference, in which the trust/distrust level between neighboring agents is portrayed as a nonlinear weight function of their opinion difference, and each agent interacts with the neighbors to update the trust/distrust level and opinion at the times determined by its own will. Based on the related properties of sub-stochastic and super-stochastic matrices, the inequality conditions about positive and negative weights to achieve opinion consensus and polarization are established. Some numerical simulations based on two well-known networks called the ``12 Angry Men" network and the Karate Club network are provided to verify the correctness of the theoretical results.
\end{abstract}


%
\IEEEpeerreviewmaketitle

\section{Introduction}\label{section:1}

In a social network, the social agents which represent the social actors, e.g., countries, parties, social individuals, will have different opinions when they are confronted with various social issues, and the opinions  may change according to the influence of other individuals.  Over the past few decades, interest in the opinion formation mechanism of agents in social networks has grown dramatically. In this regard, the DeGroot model (e.g., see \cite{DeGroot1974Consensus,DeMarzo2003Persuasion,Hamedmoghadam2019A,Hudson2020Behavior}) is a classic reference, in which each agent utilizes the opinions information from the neighbors to form its opinion on various social issues. It has been shown conclusively that when the network structure has sufficient connectivity, the social interaction in the DeGroot model tends to facilitate the agents' opinions to merge into a common opinion, namely, to reach opinion consensus in the whole social network. Based on the DeGroot model, researchers have conducted extensive researches on the mechanisms of opinion formation and constructed some improved models, such as the Friedkin-Johnsen model \cite{FJ1999Social,Ghaderi2014Opinion,Jia2015Opinion,Parsegov2017Novel,Friedkin2015The} and the Hegselmann-Krause model \cite{HK2002Opinion,Lorenz2007Continuous}. Under the existing mechanisms of opinion formation, different dynamics phenomena may emerge in the entire network, such as opinion consensus and opinion polarization, etc.

In many real situations of social networks, it is found that certain agents, called \emph{opinion leaders}, have more power on influencing the opinions, decisions and actions of most other agents (known as followers) because of their expertise or positions \cite{Roch2005The,Estrada2013How}. For instance, in the famous Karate Club network proposed in \cite{Zachary1977}, the club supervisor and the club coach had an initial conflict over the price of karate lessons. Over time, the entire club eventually was split into two new organizations led by the supervisor and the coach respectively. A study \cite{Xu2014} on the Twitter activism network found that the Twitter users with higher connectivity and problem participation, such as journalists and sports stars, had a significant impact on the opinions of ordinary Twitter users. Compared to the case in the absence of opinion leaders, it was shown in \cite{Liu2015Identi} that opinions tend to spread faster in the presence of opinion leaders. The emergence of these real social networks undoubtedly indicates that the study of the influence of opinion leaders on the formation of other agents' opinions in social networks has potential application value.  Up to now, numerous studies have focused on the dynamics of opinion consensus in the presence of opinion leaders, see \cite{Shao2018On,Dietrich2017Control,Wongkaew2015,Zhao2016Bounded,Dong2017Managing} for examples. Specifically, following findings from experimental social psychology, the reference \cite{Dietrich2017Control} considered the opinion dynamics of the DeGroot model via leadership with state and time-dependent characteristics. The opinion dynamics of the Hegselmann-Krause model with an opinion leader was examined in \cite{Wongkaew2015,Zhao2016Bounded}. To better understand the leadership mechanism, Dong \emph{et al.} \cite{Dong2017Managing} analyzed the opinion consensus of enterprises and administrations in the management field.


The aforementioned opinion dynamics models mostly considered the positive interaction relationship between agents, which is also stated as trust, cooperation, or friendliness equivalently. However, the negative interaction relationship between agents, such as distrust, competition, or antagonism, is ubiquitous in real social networks (e.g., see \cite{Wasserman1994Social,Easley2010Networks,Xue2020Opinion}). The positive and negative interaction relationships are usually represented by the edges with positive and negative weights in the signed digraph, respectively. For consistency, in this paper, the positive and negative interaction relationships are collectively referred to as ``trust" and ``distrust" respectively, and the interactive network is called the ``signed social network". A popular theory in the signed social network is structural balance \cite{David2010Networks}, in which the agents are divided into two subsets, and there is only trust between agents in the same subset and only distrust between agents belonging to different subsets. In the pioneering study \cite{Altafini2013Consensus}, Altafini proposed a simple and instructive dynamics model of signed social networks based on Laplacian feedback designs, and analyzed the dynamics of networks with and without structural balance. Since then, the Altafini model was extensively studied in \cite{Xia2016Structural,Proskurnikov2016Opinion,Liu2019Polar,Bol2020Opini}. Depending on the network structure, there will eventually be various dynamic evolution phenomena in the Altafini model, such as opinion consensus and opinion polarization. In particular, if the network is structurally balanced, the opinions of all agents gradually reach two different consensus values with the same modulus but opposite signs, namely, opinion polarization is realized. On the other hand, opinion consensus can be achieved when the agents interact by a structurally unbalanced network, where the opinions of all agents gradually converge to a common value equal to zero.

Whereas the existing Altafini model on signed social networks has been analyzed in depth, few studies have considered the influence of a leader on the evolution of opinions on the entire network. In addition, it is assumed in the Altafini model that the trust/distrust level between neighboring agents (usually represented by the weight of the edge connecting them) is independent of their own opinions. However, according to our daily life experience, it can  be observed that long-term opinion difference will lead two good friends who trust each other to drift away; in turn, holding similar opinions on certain events will weaken the gulf between two agents who distrust each other. In fact, it has been pointed out in ``Social psychology" \cite{Myers} that opinion difference is an important factor affecting the intimacy between social agents in the real interaction scenario, and people usually tend to be more trusting to the agents who have similar opinions with themselves and more distrustful towards the agents who hold different opinions with themselves.

An American classic film called ``12 Angry Men" in 1957 is a living example to show the opinion evolution of the agents under the influence of an opinion leader, in which twelve jurors were invited to decide whether a teenager was guilty of killing his father. At the beginning, only juror 8 firmly believed that the boy was acquitted, and the remaining jurors considered the boy guilty. After rounds of  discussions, juror 8 finally succeeded in persuading other eleven jurors to agree with him. Moreover, through careful observation of the details of the film, it can also be found that when juror 8 presented the evidence in support of his own opinion, other jurors showed varying attitudes (i.e., trust or distrust) to juror 8, and the trust/distrust level was affected mainly by their opinion difference with juror 8. Although juror 3, for instance, had the strongest initial distrust level to juror 8, his initial judgment was gradually shaken with jury 8's statements of sufficient evidences of the boy's innocence. With the deepening process of the opinion interactions, juror 3 gradually reduced the distrust to juror 8, and finally admitted that the existing evidence is not enough to make the guilty verdict.

Inspired by the above discussions, the main purpose of this paper is to analyze the leader-driven opinion dynamics in signed social networks with asynchronous trust/distrust level evolution. The main contributions of this paper can be summarized as follows.
\begin{enumerate}
\item [I)] An asynchronous evolution mechanism of trust/distrust level based on opinion difference is designed, where the trust/distrust level between neighboring agents is set to a weight function of their opinion difference, and each agent interacts with the neighbors to update the trust/distrust level and opinion at the times determined by its own will. To analyze the leader-driven opinion dynamics under the asynchronous evolution mechanism of trust/distrust level, we first construct the corresponding signed subgraphs to describe the asynchronous interaction scenario, and further transform the nonlinear model into a state-space form, in which the coefficient matrix is sub-stochastic or super-stochastic and its each element is a function of opinion difference between neighboring agents. In other words, the analysis of the dynamics evolution of the nonlinear model is equivalent to the analysis of the convergence of the infinite products of \emph{sub-stochastic} or \emph{super-stochastic} matrices. By developing some new approaches based on the properties of sub-stochastic and super-stochastic matrices, we solve these convergence problems, and establish mathematical inequalities of positive and negative weights to achieve consensus and polarization of opinions, respectively.

\item [III)] We design an Altafini-style interaction rule to ensure that opinion polarization can be achieved over a structurally balanced network under the asynchronous evolution of trust/distrust level. However, when using the Altafini-style interaction rule on a structurally unbalanced network, it is found that the final opinions of the followers are the convex combinations of the leader's opinion and its opposing opinion, but no consensus is reached unless the leader's opinion is preset to zero, which is theoretically confirmed in this paper. In view of this, we design another DeGroot-style interaction rule, under which the algebraic condition for achieving opinion consensus on signed social network is established in the presence of an opinion leader. Finally, it is shown that the derived results of opinion polarization and opinion consensus can be applied to explain the opinion evolution phenomena of real-world networks in social psychology, such as the ``12 Angry Men" network and the Karate Club network.
\end{enumerate}

The rest of this paper is arranged as follows. Section~\ref{section:2} introduces some basic notations and preliminaries about the interaction network and matrix. The problem statement is shown in Section~\ref{section:3}. Section~\ref{section:4} and Section~\ref{section:5} present the main results on opinion polarization and opinion consensus, respectively. Section~\ref{section:6} validates the correctness of the theoretical results by some simulation examples. The paper is concluded in Section~\ref{section:7}.

\section{Preliminaries}\label{section:2}

\subsection{Matrix notations}\label{section:2.1}

Some notations for a real matrix $S\!=\![s_{ij}]_{n\times n}$ are introduced as follows. $\diag\{S\}$ denotes a diagonal matrix containing sequential diagonal elements $s_{11},s_{22},\ldots,s_{nn}$. $|S|=[|s_{ij}|]_{n\times n}$ stands for a nonnegative matrix in which each element $|s_{ij}|$ is the absolute value of $s_{ij}$. The $i$th row's sum of matrix $S$ is represented by $\Lambda_i[S]=\sum_{j=1}^{n}s_{ij}$. The infinite norm of matrix $S$ is expressed as $\|S\|_{\infty}=\max\big\{\Lambda_i[|S|]\mid i=1,2,\ldots,n\big\}$. The real matrix $S$ is nonnegative (non-positive) if $s_{ij}\geq0$ (if $s_{ij}\leq0$) for any $i,j=1,2,\ldots,n$. Let $S(i:j,p:q)$ be a block matrix consisting of the $i$-th row to the $j$-th row and the $p$-th column to the $q$-th column of matrix $S$, where $i\leq j, p\leq q$. For convenience, $[S]_{ij}$ can also be used to denote the element $s_{ij}$ in $S$ if there is no ambiguity. Denote the left products of matrices $S_i, i=1,2,\ldots,q$ by  $\prod_{t=1}^qS_t=S_q\cdots S_2S_1$. $\mathbf{0}$ is a matrix with all elements being 0. $|\mathscr{N}|$ represents the number of elements in the set $\mathscr{N}$.

\subsection{Supporting definitions and lemmas}\label{section:2.2}

\begin{definition}\cite{Chen2013Consensus}\label{definition:1}
A real matrix $S=[s_{ij}]_{n\times n}$ is \emph{general row-stochastic} if $\Lambda _i[S]=1$, $i=1,2,\ldots,n$.
\end{definition}

\begin{definition}\cite{Pullman1966Infinite}\label{definition:2}
A real matrix $S=[s_{ij}]_{n\times n}$ is \emph{sub-stochastic} if it is nonnegative and $\Lambda _i[S]\leq1$, $i=1,2,\ldots,n$.
\end{definition}

\begin{definition}\label{definition:3}
A real matrix $S=[s_{ij}]_{n\times n}$ is \emph{super-stochastic} if it is nonnegative and there exists a set $\mathcal{W}\subset\{1,2,\ldots,n\}$ such that $\Lambda_{i}[S]<1$, $i\in\mathcal{W}$ and $\Lambda_{i}[S]\geq1$, $i\notin\mathcal{W}$.
\end{definition}

\begin{lemma}\cite{Shi2020Sub}\label{lemma:1}
Let $S_1,S_2,\ldots,S_q$ be $n\times n$ super-stochastic matrices and $\max\{\Lambda_{i}[S_t]\mid t=1,2,\ldots,q, \ i=1,2,\ldots,n\}=g$, then $\|\prod_{t=1}^qS_t\|_{\infty}\leq g^q$.
\end{lemma}

Similar to Lemma \ref{lemma:1}, we present the corresponding results for the general row-stochastic matrix and the sub-stochastic matrix in the following lemmas, respectively.

\begin{lemma}\label{lemma:2}
Let $S_1,S_2,\ldots,S_q$ be $n\times n$ general row-stochastic matrices, then $\prod_{t=1}^qS_t$ is a general row-stochastic matrix.
\end{lemma}

\begin{lemma}\label{lemma:3}
Let $S_1,S_2,\ldots,S_q$ be $n\times n$ sub-stochastic matrices, then $\prod_{t=1}^qS_t$ is a sub-stochastic matrix.
\end{lemma}


\subsection{Signed social network}\label{section:2.3}

Consider a social network with $n$ agents indexed in the set $\mathscr{V}=\{v_1,v_2,\ldots,v_n\}$, where $v_n$ represents the opinion leader and $v_1,v_2,\ldots,v_{n-1}$ stand for the followers. Each agent $v_i$ holds an opinion $x_i\in\mathbb{R}$. The opinion leader is an agent with a firm stand and its opinion is not affected by the opinions of other agents. The opinions of followers are influenced by those of their neighbors.

The structure of the social network is represented as a signed digraph $\mathscr{G}=(\mathscr{V},\mathscr{E})$, where $\mathscr{V}$ and $\mathscr{E}\subseteq\mathscr{V}\times\mathscr{V}$ represent the sets of nodes and edges, respectively. There is a directed edge $(v_j,v_i)$ in $\mathscr{E}$ if and only if agent $v_i$ takes agent $v_j$ as an \emph{in-neighbor} and thus agent $v_j$'s opinion is influencing agent $v_i$'s. Denote the signed adjacency matrix by $\mathscr{A}=[a_{ij}]$ with the elements being 0, 1 or $-1$, where $a_{ij}\neq 0$ if and only if $(v_j,v_i)\in\mathscr{E}$, otherwise $a_{ij}=0$. In particular, $a_{ij}=1$ and $a_{ij}=-1$ represent that agent $v_i$ trust and distrust agent $v_j$, respectively. The sets of trusted \emph{in-neighbors} and distrustful \emph{in-neighbors} of agent $v_i$ are denoted by  $\mathscr{N}^+_i=\{v_j\mid(v_j,v_i)\in\mathscr{E}, a_{ij}=1\}$ and $\mathscr{N}^-_i=\{v_j\mid(v_j,v_i)\in\mathscr{E}, a_{ij}=-1\}$, respectively.
The set of in-neighbors of agent $v_i$ is $\mathscr{N}_i=\mathscr{N}^+_i\cup\mathscr{N}^-_i$. For convenience, the agent's neighbors refer specifically to its in-neighbors in the rest of this paper.

A directed path from node $v_{i_1}$ to node $v_{i_z}$ in a digraph $\mathscr{G}$ is denoted by $\mathscr{P}_{v_{i_1}\rightarrow v_{i_z}}=v_{i_1}\rightarrow v_{i_2}\rightarrow \cdots\rightarrow v_{i_z}$, where $v_{i_1},v_{i_2},\ldots,v_{i_z}$ are distinct and $(v_{i_s-1},v_{i_s})\in\mathscr{E}$ for $s=2,3,\ldots,z$. The directed distance from $v_{i_1}$ to $v_{i_z}$ is the number of edges in the shortest path from $v_{i_1}$ to $v_{i_z}$. A digraph $\mathscr{G}$ is strongly connected if there is a path from $v_i$ to $v_j$ between any pair of distinct nodes $v_i, v_j$, and it is weakly connected if replacing all of its directed edges with undirected edges produces a undirected connected graph. A digraph $\mathscr{G}_1$ is called a strongly connected component of $\mathscr{G}$ if it is a maximal strongly connected subgraph of $\mathscr{G}$, and further, $\mathscr{G}_1$ is a closed strongly connected component of $\mathscr{G}$ if there are no edges reaching the nodes in $\mathscr{G}_1$. In a weakly connected digraph, for any node $v_i$ outside the closed strongly connected components, there exists at least a path from the closed strongly connected components to it. In addition, for $\mathscr{V}^*\subseteq\mathscr{V}$, let $\mathscr{G}(\mathscr{V}^*)$ represent $\mathscr{G}$'s an induced subgraph with the edge set $\mathscr{E}(\mathscr{V}^*)$, where $(v_i,v_j)\in\mathscr{E}(\mathscr{V}^*)$ if $v_i,v_j\in\mathscr{V}^*$ and $(v_i,v_j)\in\mathscr{E}$.

Structural balance is a basic concept in the study of signed graph, and its main motivation is social interpersonal and economic networks. The definition of structural balance is given as follows.

\begin{definition}\cite{David2010Networks}\label{definition:4}
A signed digraph $\mathscr{G}$ is structurally balanced if there is a partition of the node set into $\mathscr{V}=\mathscr{V}_1\cup\mathscr{V}_2$ with $\mathscr{V}_1$, $\mathscr{V}_2$ being non-empty and mutually disjoint, where for any $(v_j,v_i)\in\mathscr{E}$, $a_{ij}=1$ if and only if $v_i$ and $v_j$ belong to the same subset, otherwise, $a_{ij}=-1$.
\end{definition}

\subsection{Asynchronous interaction}\label{section:2.4}

\begin{figure}[h]
  \centering
    \includegraphics[width=8cm,height=3cm]{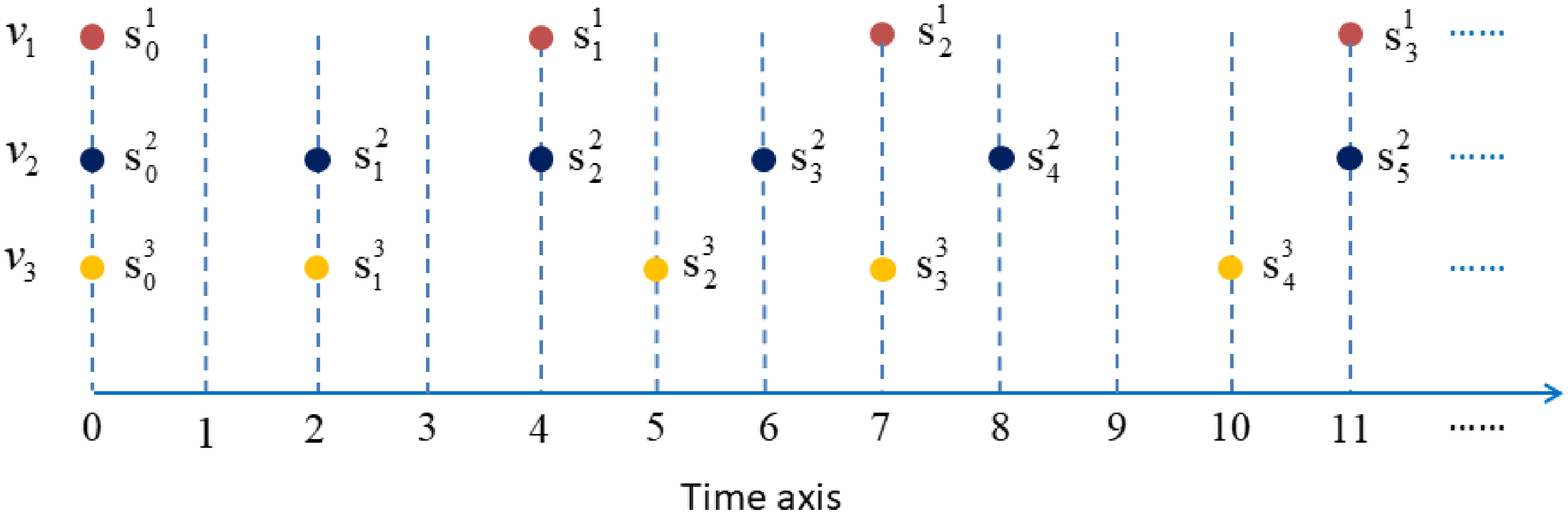}
  \caption{An example of interaction times for agents $v_1$, $v_2$ and $v_3$ in the asynchronous interaction scenario.}
  \label{fig1}
 \end{figure}

Most of the existing literature is dedicated to analyzing opinion dynamics of social networks under the synchronous interaction, in which all agents interact with their neighbors at the same time. However, in real-world social networks, the synchronous interaction is generally not easy to implement due to some objective factors and the subjective will. This consideration shifts our attention to a more general asynchronous interaction scenario in this paper, where each agent interacts with the neighbors to update the trust/distrust level and opinion at the times determined by its own will. In other words, each agent can interact with its neighbors at arbitrary times in the asynchronous interaction scenario. Without loss of generality, we assume that the set of each agent $v_i$'s opinion interaction times is
\begin{equation*}
\begin{aligned}
\{s^{i}_{k}\}=\{s^{i}_{0},s^{i}_{1},\ldots,s^{i}_{k},\ldots\},
\end{aligned}
\end{equation*}
in which $s^{i}_{0},s^{i}_{1},\ldots,s^{i}_{k}\in\{0,1,\ldots,t,\ldots\}$ and $0=s^{i}_{0}<s^{i}_{1}<\cdots<s^{i}_{k}<\cdots$. It is further assumed that $\{s^{i}_{k}\}$ satisfies the following condition:
\begin{eqnarray}\label{eq:1}
s^{i}_{k+1}-s^{i}_{k}\leq h,
\end{eqnarray}
where $h$ is a positive integer. An example for three agents' interaction times in the asynchronous interaction scenario is shown in Fig. \ref{fig1}.

\begin{remark}\label{remark:1}
Condition (\ref{eq:1}) is given to guarantee that there is no such an agent, who no longer communicates with its neighbors from a certain time. Equivalently, in each finite time interval with length $h$, where $h$ is a constant that can be arbitrarily large, each agent interacts with its neighbors at least once. This is necessary in the asynchronous interaction scenario. Without Condition (\ref{eq:1}), there may be an agent unable to interact with its neighbors at all times, which obviously cannot guarantee the realization of opinion consensus or opinion polarization. In addition, it is worth noting that the asynchronous interaction considered in this paper includes the synchronous interaction as a special case of $h=1$. Therefore, the asynchronous dynamics results obtained in this paper can be easily extended to the synchronous interaction scenario.
\end{remark}

\section{Model formulation}\label{section:3}

In a social network, the trust/distrust level between neighbouring agents is usually affected by their opinion difference. For example, it can be observed in the daily interaction that two closely related agents will gradually alienate due to long-term opinion difference, and the similarity of opinions will gradually eliminate the gulf between two distrusting agents. Take into account such an interaction scenario, we set the trust/distrust level of agent $v_i$ to agent $v_j$ as a function $f_{ij}$ of their opinion difference $|x_j(t)-x_i(t)|$, i.e.,
\begin{equation}\label{eq:2}
f_{ij}(t)\!=\!
\left\{\!
\begin{aligned}
&f^+_{ij}\big(|x_j[t]\!-\!x_i[t]|\big), \ {\rm if } \ a_{ij}\!=\!1,\\
&f^-_{ij}\big(|x_j[t]\!-\!x_i[t]|\big), \ {\rm if } \ a_{ij}\!=\!-1,
\end{aligned}
\right.
\end{equation}
where the weight functions $f^+_{ij}$ and $f^-_{ij}$ are used to quantify the trust and distrust levels of agent $v_i$ to agent $v_j$, respectively. Suppose the weight functions satisfy the following assumptions.

\begin{assumption}\label{assumption:1}
For each agent $v_i$, the functions $f^+_{ij}$, $j=1,2,\ldots,n$ are positive, decreasing and bounded, in particular, $\underline{\alpha}\!\leq \! f^+_{ij}(y_2)\!\leq\! f^+_{ij}(y_1)\!\leq\!\overline{\alpha}$, $j\!\neq \!n$ and $\underline{\beta}\!\leq\! f^+_{in}(y_2)\!\leq \! f^+_{in}(y_1)\!\leq\!\overline{\beta}$ as $0\!\leq\! y_1\!\leq\! y_2\!<\!+\infty$.
\end{assumption}

\begin{assumption}\label{assumption:2}
For each agent $v_i$, the functions $f^-_{ij}$, $j=1,2,\ldots,n$ are positive, increasing and bounded, in particular, $\underline{\iota}\!\leq \! f^-_{ij}(y_1)\!\leq \!f^-_{ij}(y_2)\!\leq\!\overline{\iota}$, $j\!\neq \!n$ and $\underline{\kappa}\!\leq\! f^-_{in}(y_1)\!\leq\! f^-_{in}(y_2)\!\leq\!\overline{\kappa}$ as $0\!\leq\! y_1\!\leq\! y_2\!<\!+\infty$.
\end{assumption}

\begin{remark}\label{remark:2}
For each edge $(v_j,v_i)$, there is a corresponding weight function $f_{ij}$ to quantify the trust/distrust level of agent $v_i$ to agent $v_j$, which reflects the difference of opinion interaction between different neighboring agents in the social network. In addition, the leader which plays a leading role in the evolution of the entire social network is usually an agent with significant influence. Therefore, for any follower with the leader as one of its neighbors, the leader's opinion often has more influence on its opinion than those of other neighbors. In order to better reflect this situation, we set different bounds for the weight functions $f^+_{ij}$, $j\!\neq \!n$ and $f^+_{in}$ in Assumption \ref{assumption:1}. The setting of different bounds is also applied to the weight functions $f^-_{ij}$, $j\!\neq \!n$ and $f^-_{in}$ in Assumption \ref{assumption:2}.
\end{remark}

Each agent $v_i$ updates its opinion in accordance with the following asynchronous opinion dynamics model
\begin{equation}\label{eq:3}
x_{i}[t+1]=\left\{
\begin{aligned}
&\theta_ix_i[t]\!+\!(1\!-\!\theta_i)u_i[t], \ t\in\{s_k^i\},\\
&x_i[t], \ \ \ \ \ \ \ \ \ \ \ \ \ \ \ \ \ \ \ \ \!t\notin\{s_k^i\},\\
\end{aligned}
\right.
\end{equation}
where $x_i[t]$ represents the opinion of agent $v_i$ at time $t$, $\theta_i\in[0,1)$ is the level of self-confidence of agent $v_i$ in its own opinion evolution, and suppose $\theta_n=1$. The interaction rule is designed as follows:
\begin{equation}\label{eq:4}
u_i[t]=\sum_{v_j\in\mathscr{N}_i}\frac{a_{ij}f_{ij}(t)}{\sum_{v_j\in\mathscr{N}_i}f_{ij}(t)}x_j[t].
\end{equation}

\begin{figure}[h]
  \centering
  \subfigure[$\mathscr{G}$]{
    \includegraphics[width=0.7in]{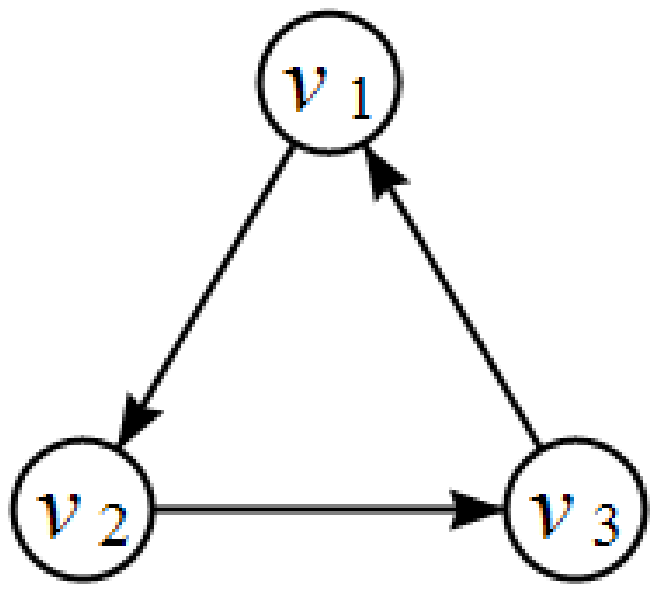}}
  \subfigure[$\mathscr{G}(0)$]{
    \includegraphics[width=0.7in]{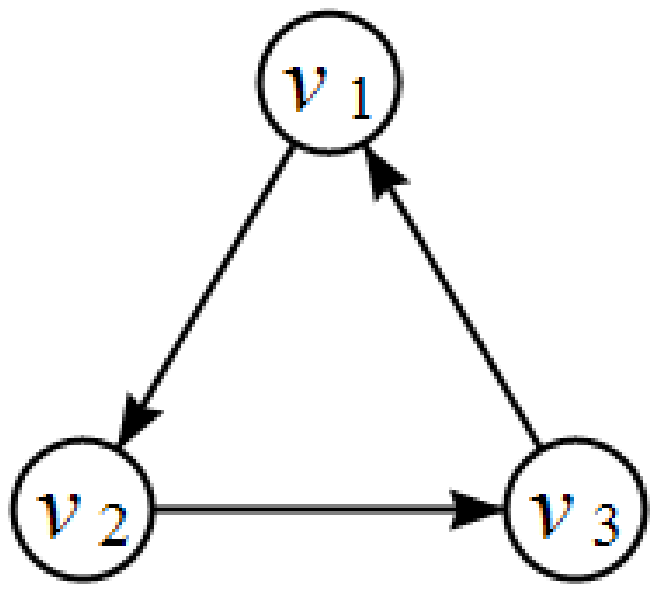}}
  \subfigure[$\mathscr{G}(1)$]{
    \includegraphics[width=0.7in]{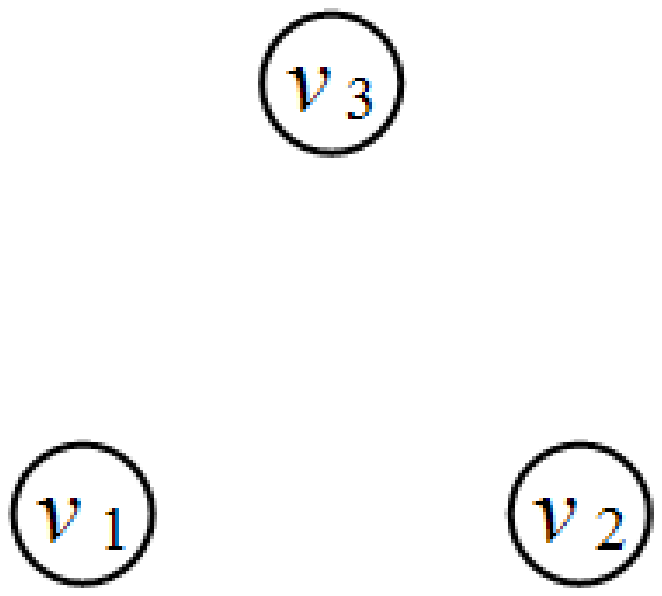}}
  \subfigure[$\mathscr{G}(2)$]{
    \includegraphics[width=0.7in]{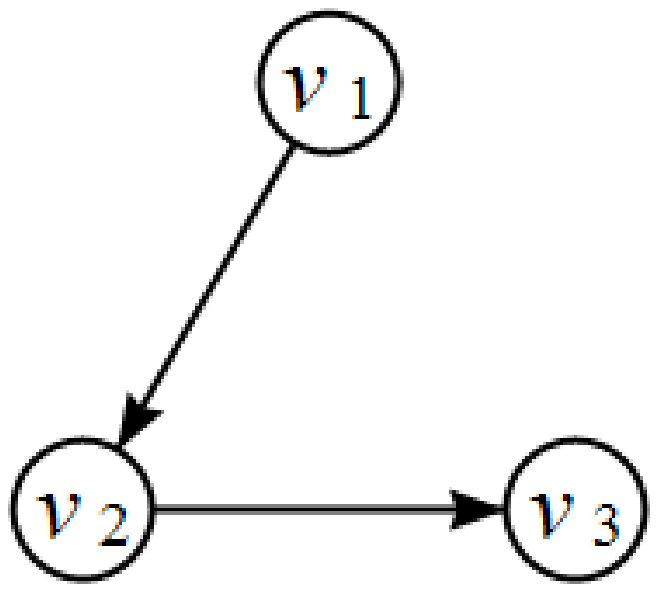}}
  \subfigure[$\mathscr{G}(3)$]{
    \includegraphics[width=0.7in]{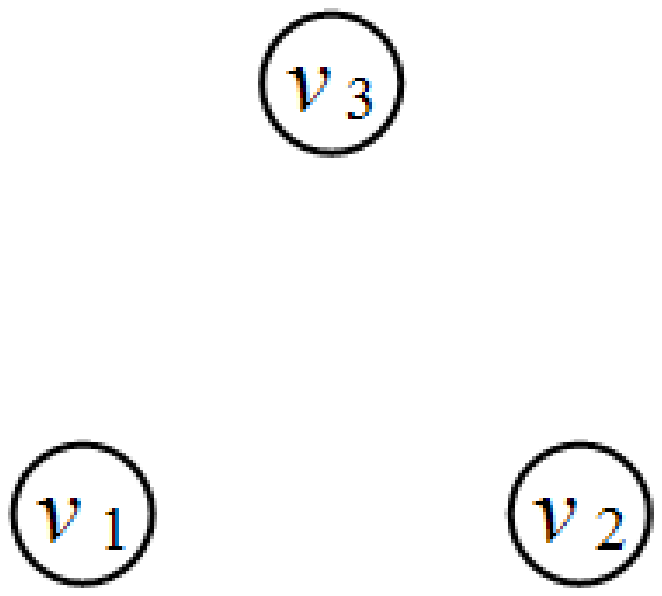}}
  \subfigure[$\mathscr{G}(4)$]{
    \includegraphics[width=0.7in]{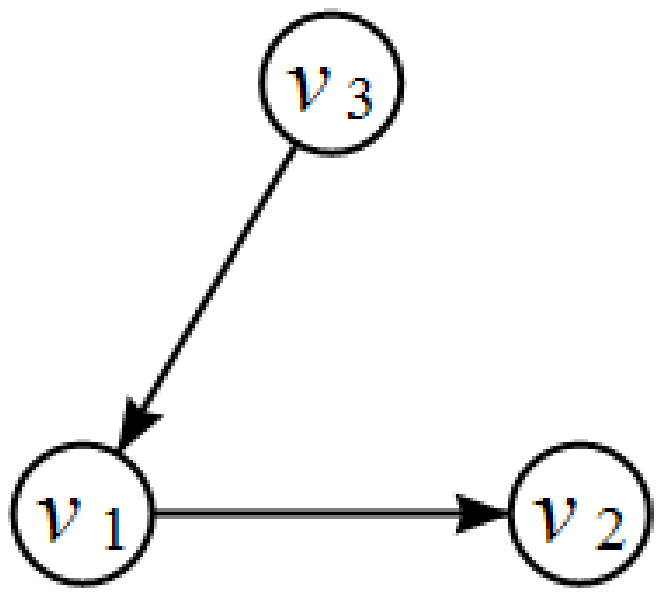}}
    \subfigure[$\mathscr{G}(5)$]{
    \includegraphics[width=0.7in]{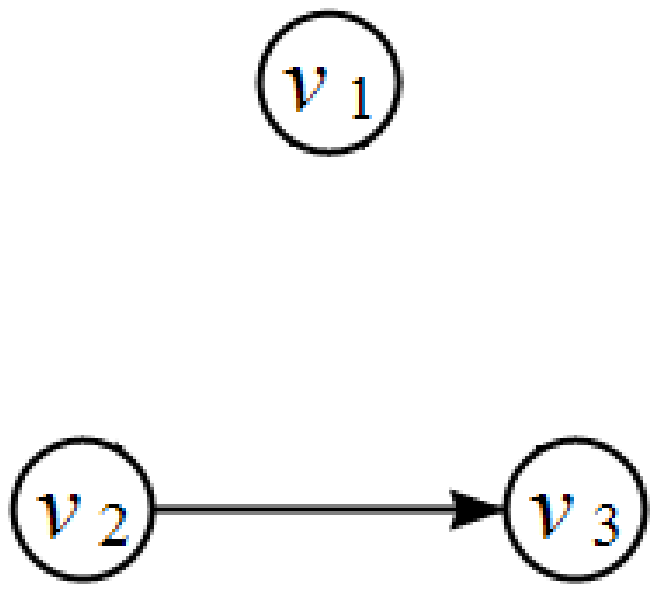}}
  \subfigure[$\mathscr{G}(6)$]{
    \includegraphics[width=0.7in]{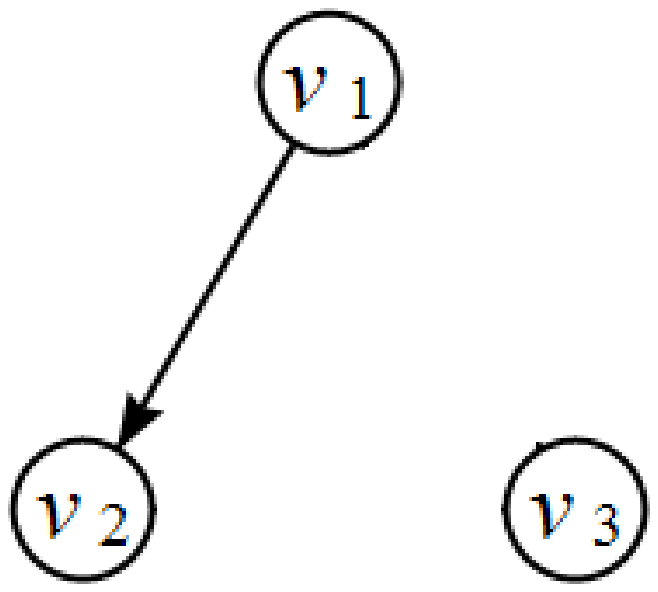}}
  \caption{An given interaction network $\mathscr{G}$ and the corresponding newly constructed signed digraphs $\mathscr{G}(t)$, $t=0,1,\ldots,6$ under the asynchronous interaction shown in Fig. \ref{fig1}.}\label{fig2}
\end{figure}

In general, it is not easy to directly analyze the evolution of agents' opinions due to the complexity of the nonlinear term $f_{ij}(t)$ in the interaction rule (\ref{eq:4}). On this account, we consider the asynchronous interaction scenario from a new perspective on the interaction network. Since all time sequences $\{s^{1}_{k}\}$, $\{s^{2}_{k}\},\ldots,\{s^{n}_{k}\}$ are independent of each other, it is uncertain which agents update their opinions at time $t$ in view of the distributed interaction. To proceed, we construct a signed digraph $\mathscr{G}(t)\!=\!\big(\mathscr{V},\mathscr{E}(t)\big)\!\subseteq\!\mathscr{G}$ for each time $t$. The signed adjacency matrix of $\mathscr{G}(t)$ is denoted by $\mathscr{A}(t)\!=\![a_{ij}(t)]$ which satisfies: $(v_j,v_i)\!\in\!\mathscr{E}(t)$ and $a_{ij}(t)\!=\!a_{ij}$, $j\!=\!1,2,\ldots,n$ if $t\!\in\!\{s_k^i\}$, and $(v_j,v_i)\!\notin\!\mathscr{E}(t)$ and $a_{ij}(t)\!=\!0$, $j\!=\!1,2,\ldots,n$ if $t\!\notin\!\{s_k^i\}$. The set of neighbors of agent $v_i$ at time $t$ is denoted by $\mathscr{N}_i(t)$. For given interaction network $\mathscr{G}$ in Fig. \ref{fig2}(a), the newly constructed signed digraphs $\mathscr{G}(t)$ at different times under the asynchronous interaction setting in Fig. \ref{fig1} are shown in Fig. \ref{fig2}(b)-\ref{fig2}(h).

\section{Discussion on opinion polarization}\label{section:4}

Distrust between agents is ubiquitous in real-world social networks. Opinion polarization is a dynamic characteristic formed by agents in signed social networks through opinion interactions, and it has received extensive attention in the Altafini model \cite{Altafini2013Consensus,Xia2016Structural,Proskurnikov2016Opinion,Liu2019Polar,Bol2020Opini}. In these studies, however, the effect of opinion difference between neighboring agents on their trust/distrust level was not considered. In this section, our purpose is to explore the algebraic conditions on the weight functions and the network structure to achieve opinion polarization under the asynchronous evolution mechanism of trust/distrust level based on opinion difference.

Consider the opinion interaction rule (\ref{eq:4}). According to the construction of signed digraph $\mathscr{G}(t)$, model (\ref{eq:3}) is equivalently expressed as
\begin{equation}\label{eq:5}
\begin{aligned}
x_i[t\!+\!1]=\theta_i(t)x_i[t]\!+\!\big(1\!-\!\theta_i(t)\big)\sum_{v_j\in\mathscr{N}_i(t)}q_{ij}(t)x_j[t],
\end{aligned}
\end{equation}
where
\begin{equation}\label{eq:6}
\begin{aligned}
q_{ij}(t)=\frac{a_{ij}(t)f_{ij}(t)}{\sum_{v_j\in\mathscr{N}_i(t)}f_{ij}(t)},
\end{aligned}
\end{equation}
and
\begin{equation}\label{eq:7}
\begin{aligned}
\left\{
\begin{aligned}
&\theta_i(t)=\theta_i, \ t\in\{s_k^i\},\\
&\theta_i(t)=1, \ \ t\notin\{s_k^i\}.
\end{aligned}
\right.
\end{aligned}
\end{equation}
We now present the sufficient condition for achieving opinion polarization.

\begin{theorem}\label{theorem:1}
Consider model (\ref{eq:3}) with the interaction rule (\ref{eq:4}). Opinion polarization is achieved if the signed digraph $\mathscr{G}$ is structurally balanced as well as contains a spanning tree rooted at the leader and the agents' self-confidence levels satisfy $\theta_i\neq0$, $i=1,2,\ldots,n-1$. In particular, $\lim_{t\rightarrow\infty}x_i[t]\!=\!x_n[0]$ if $v_i$ and $v_n$ are in the same subset, and $\lim_{t\rightarrow\infty}x_i[t]\!=\!-x_n[0]$ if $v_i$ and $v_n$ belong to different subsets.
\end{theorem}

\begin{IEEEproof}
Since the signed digraph $\mathscr{G}$ is structurally balanced, there is a partition of $\mathscr{V}$ into $\mathscr{V}=\mathscr{V}_1\cup\mathscr{V}_2$. Without loss of generality, we assume $\mathscr{V}_1=\{v_1,v_2\ldots,v_m\}$ and $\mathscr{V}_2=\{v_{m+1},v_{m+2},\ldots,v_n\}$. Then the matrix $Q(t)=[q_{ij}(t)]$ takes a block-matrix form.
\begin{equation*}
Q(t)=\begin{bmatrix}
       Q_{11}(t) & Q_{12}(t) \\
       Q_{21}(t) & Q_{22}(t)
     \end{bmatrix},
\end{equation*}
where $Q_{11}(t),Q_{22}(t)\geq0$ and $Q_{12}(t),Q_{21}(t)\leq0$. Note that $|Q(t)|$ is a row-stochastic matrix. Let
\begin{equation*}
x^*[t]=\big[\!-\!x_1[t],\ldots,\!-\!x_m[t],x_{m+1}[t],\ldots,x_{n}[t]\big]^T,
\end{equation*}
then we can rewrite Eq. (\ref{eq:5}) into a state-space form
\begin{equation}\label{eq:8}
x^*[t+1]=\Omega(t) x^*[t],
\end{equation}
where $\Xi[t]$ is a time-varying row-stochastic matrix with the following details
\begin{equation*}
\Omega(t)=\begin{bmatrix}
              \Psi(t) & r(t) \\
              \mathbf{0} & 1 \\
            \end{bmatrix},
\end{equation*}
in which
\begin{equation*}
\begin{aligned}
&\Psi(t)\!=\!\Theta^*(t)\!+\!\big(I_{n-1}-\Theta^*(t)\big)|Q(t)|(1:n\!-\!1,1:n\!-\!1),\\
&r(t)\!=\!\big(I_{n-1}\!-\!\Theta^*(t)\big)|Q(t)|(1:n\!-\!1,n:n),\\
&\Theta^*(t)\!=\!{\rm diag}\big\{\theta_1(t),\theta_2(t),\!\ldots\!,\theta_{n-1}(t)\big\}.
\end{aligned}
\end{equation*}
Under the condition $\theta_i\neq0$, $\Psi(t)$ is a sub-stochastic matrix with positive diagonal elements, and its each non-diagonal element is a multivariate function, i.e.,
\begin{equation*}
\begin{aligned}
\big[\Psi(t)\big]_{ij}=g\Big(a_{ij}(t),\theta_{i}(t),f_{ij}(t)\Big), \ i\neq j,
\end{aligned}
\end{equation*}
where
\begin{equation*}
\begin{aligned}
&g\Big(a_{ij}(t),\theta_{i}(t),f_{ij}(t)\Big)\in[\epsilon,1), \ {\rm if} \ a_{ij}(t)\neq0,\\
&g\Big(a_{ij}(t),\theta_{i}(t),f_{ij}(t)\Big)=0, \ \ \ \ \ {\rm if} \ a_{ij}(t)=0,
\end{aligned}
\end{equation*}
in which
\begin{equation*}
\begin{aligned}
\epsilon=\min\left\{\theta_{min},\frac{(1-\theta_{max})\min\{\underline{\alpha},\underline{\beta},\underline{\iota},\underline{\kappa}\}}
{\max\{\overline{\alpha},\overline{\beta},\overline{\iota},\overline{\kappa}\}|\mathscr{N}_{max}|}\right\}
\end{aligned}
\end{equation*}
with
\begin{equation*}
\begin{aligned}
&\theta_{min}=\min\{\theta_i \mid i=1,2,\ldots,n-1\},\\
&\theta_{max}=\max\{\theta_i \mid i=1,2,\ldots,n-1\},\\
&|\mathscr{N}_{max}|=\max\{|\mathscr{N}_{i}| \mid i=1,2,\ldots,n-1\}.
\end{aligned}
\end{equation*}
Furthermore, the row sums of $\Psi(t)$ satisfy: if $t\in\{s_k^i\}$ and $(v_n,v_i)\in\mathscr{E}$, that is $a_{ij}(t)\neq0$,  then
\begin{equation*}
\begin{aligned}
\Lambda_i[\Psi(t)]&=1-g\Big(a_{in}(t),\theta_{i}(t),f_{in}(t)\Big)\\
&\leq1-\epsilon<1,
\end{aligned}
\end{equation*}
and if $t\in\{s_k^i\}$ and $(v_n,v_i)\!\notin\!\mathscr{E}$, or $t\!\notin\!\{s_k^i\}$, namely, $a_{ij}(t)=0$, then $\Lambda_i[\Psi(t)]\!=\!1$.

To achieve opinion polarization for model (\ref{eq:5}), the following equation needs to be proved first
\begin{equation}\label{eq:9}
\prod_{t=0}^{\infty}\Psi(t)=\mathbf{0}.
\end{equation}
For this purpose, the discrete times $0,1,\ldots,t,\ldots$ are divided into a series of intervals $[\gamma ph,\gamma ph+ph)$, $\gamma\in\mathbb{N}$, where $p$ is the longest distance from the leader to the followers on the spanning tree of the signed digraph $\mathscr{G}$. Then, we analyze the matrices product $\prod_{t=\gamma ph}^{\gamma ph+ph-1}\Psi(t)$ associated with each interval $[\gamma ph,\gamma ph+ph)$ through the following two parts.

\emph{Part I:} Consider the matrices product $\prod_{t=\gamma ph}^{\gamma ph+h-1}\Psi(t)$ associated with the interval $[\gamma ph,\gamma ph+h)$. For each follower $v_{s_{z}}$, there exists a path from the leader $v_n$ to it since the signed digraph $\mathscr{G}$ contains a spanning tree rooted at the leader. Suppose that the path from $v_n$ to $v_{s_{z}}$ is described as $\mathscr{P}_{v_n\rightarrow v_{s_z}}=v_n\rightarrow v_{s_{1}}\rightarrow\cdots\rightarrow v_{s_{z}}$. It is known from condition (\ref{eq:1}) that each agent updates its opinion at least once in arbitrary time interval with length $h$. Thus, there exists a time $\gamma ph+l_1$ such that $\gamma ph\leq\gamma ph+l_1<\gamma ph+h$ and $(v_n,v_{s_1})\in\mathscr{E}(\gamma ph+l_1)$, which follows that $\Lambda_{s_1}[ \Psi(\gamma ph+l_1)]\leq1-\epsilon$. Noting the result derived by Lemma \ref{lemma:3} that
\begin{small}
\begin{equation*}
\begin{aligned}
&\Lambda_{i}\Big[ \prod_{t=\gamma ph}^{\gamma ph+l_1-1}\Psi(t)\Big]\leq 1, \ \Lambda_{i}\Big[\prod_{t=\gamma ph+l_1+1}^{\gamma ph+h-1}\Psi(t)\Big]\leq 1
\end{aligned}
\end{equation*}
\end{small}

\noindent
for $i=1,2,\ldots,n-1$, there holds
\begin{small}
\begin{equation*}
\begin{aligned}
\Lambda_{s_1}\Big[\prod_{t=\gamma ph}^{\gamma ph+l_1}\!\!\Psi(t)\Big]&=\sum_{j}\big[\Psi(\gamma ph\!+\!l_1)\big]_{s_1j}\Lambda_{j}\Big[\prod_{t=\gamma ph}^{\gamma ph+l_1-1}\!\!\Psi(t)\Big]\\
& \leq\Lambda_{s_1}\big[\Psi(\gamma ph\!+\!l_1)\big]\leq1-\epsilon.
\end{aligned}
\end{equation*}
\end{small}

\noindent
This combines the fact $\big[\prod_{t=\gamma ph+l_1+1}^{\gamma ph+h-1}\Psi(t)\big]_{s_1s_1}\!\geq\!\epsilon^{h-l_1-1}$ to guarantee that
\begin{small}
\begin{equation}\label{eq:10}
\begin{aligned}
\Lambda_{s_1}\Big[\prod_{t=\gamma ph}^{\gamma ph+h-1}\!\!\Psi(t)\Big]&\!=\!\sum_{j\neq s_1}\!\Big[\prod_{t=\gamma ph+l_1+1}^{\gamma ph+h-1}\!\!\Psi(t)\Big]_{s_1j}\Lambda_{j}\Big[\prod_{t=\gamma ph}^{\gamma ph+l_1}\!\!\Psi(t)\Big]\\
& \ \ +\Big[\prod_{t=\gamma ph+l_1+1}^{\gamma ph+h-1}\!\!\Psi(t)\Big]_{s_1s_1}\Lambda_{s_1}\Big[\prod_{t=\gamma ph}^{\gamma ph+l_1}\!\!\Psi(t)\Big]\\
&\leq(1\!-\!\epsilon^{h-l_1-1})\!+\!\epsilon^{h-l_1-1}(1\!-\!\epsilon)\\
&\leq1\!-\!\epsilon^{h}.
\end{aligned}
\end{equation}
\end{small}

\emph{Part II:} Consider the matrices product $\prod_{t=\gamma ph+h}^{\gamma ph+ph-1}\Psi(t)$ associated with the interval $[\gamma ph+h,\gamma ph+ph)$. Since each agent $v_{s_{y-1}}$ updates its opinion at least once in the interval $[\gamma ph+yh-h,\gamma ph+yh)$, where $y\in\{2,3,\ldots,z\}$, there exists a time $\gamma ph+yh-h+l_y$ such that $(v_{s_{y-1}},v_{s_{y}})\in\mathscr{E}(\gamma ph+yh-h+l_y)$, where $0\leq l_y<h$. This implies that $[\Psi(\gamma ph+yh-h+l_y)]_{s_ys_{y-1}}\geq\epsilon$, which along with the facts
\begin{small}
\begin{equation*}
\begin{aligned}
&\Big[\prod_{t=\gamma ph+yh-h}^{\gamma ph+yh-h+l_y-1}\Psi(t)\Big]_{s_{y-1}s_{y-1}}\geq\epsilon^{l_y},\\
&\Big[\prod_{t=\gamma ph+yh-h+l_y+1}^{\gamma ph+yh-1}\Psi(t)\Big]_{s_{y}s_{y}}\geq\epsilon^{h-l_y-1},
\end{aligned}
\end{equation*}
\end{small}

\vspace{-1ex}
\noindent
verifies that
\begin{small}
\begin{equation}\label{eq:11}
\begin{aligned}
\Big[\prod_{t=\gamma ph+yh-h}^{\gamma ph+yh-1}\!\!\Psi(t)\Big]_{s_{y}s_{y-1}}\!\!\geq&\Big[\prod_{t=\gamma ph+yh-h+l_y+1}^{\gamma ph+yh-1}\Psi(t)\Big]_{s_{y}s_{y}}\\
&\times[\Psi(\gamma ph+yh-h+l_y)]_{s_ys_{y-1}}\\
&\times\!\Big[\!\prod_{t=\gamma ph+yh-h}^{\gamma ph+yh-h+l_y-1}\!\!\Psi(t)\Big]_{s_{y-1}s_{y-1}}\!\!\geq\!\!\epsilon^h.
\end{aligned}
\end{equation}
\end{small}

\noindent
It follows that
\begin{small}
\begin{equation}\label{eq:12}
\begin{aligned}
\Big[\prod_{t=\gamma ph+h}^{\gamma ph+zh-1}\!\!\Psi(t)\Big]_{s_{z}s_{1}}&\geq\prod_{y=2}^z\Big[\prod_{t=\gamma ph+yh-h}^{\gamma ph+yh-1}\!\!\Psi(t)\Big]_{s_{y}s_{y-1}}\\
&\geq\epsilon^{(z-1)h}.
\end{aligned}
\end{equation}
\end{small}

\noindent
Since $p\!\geq\! z$ and $\big[\prod_{t=\gamma ph+zh}^{\gamma ph+ph-1}\Psi(t)\big]_{s_{z}s_{z}}\!\geq\!\epsilon^{(p-z-1)h}$, we have
\begin{small}
\begin{equation}\label{eq:13}
\begin{aligned}
\Big[\!\!\prod_{t=\gamma ph+h}^{\gamma ph+ph-1}\!\!\!\Psi(t)\!\Big]_{s_{z}s_{1}}&\!\!\!\geq\!\Big[\!\!\prod_{t=\gamma ph+zh}^{\gamma ph+ph-1}\!\!\!\Psi(t)\Big]_{s_{z}s_{z}}
\Big[\!\!\prod_{t=\gamma ph+h}^{\gamma ph+zh-1}\!\!\!\Psi(t)\Big]_{s_{z}s_{1}}\\
&\!\!\!\geq\epsilon^{(p-1)h}.
\end{aligned}
\end{equation}
\end{small}

\vspace{-2ex}
Combining (\ref{eq:10}) and (\ref{eq:13}), we get
\begin{small}
\begin{equation*}
\begin{aligned}
\Lambda_{s_z}\Big[\prod_{t=\gamma ph}^{\gamma ph+ph-1}\!\!\Psi(t)\Big]&=\sum_{j\neq s_1}\!\Big[\prod_{t=\gamma ph+h}^{\gamma ph+ph-1}\!\!\Psi(t)\Big]_{s_{z}j}\Lambda_{j}\Big[\prod_{t=\gamma ph}^{\gamma ph+h-1}\Psi(t)\Big]\\
& \ \ \ +\Big[\prod_{t=\gamma ph+h}^{\gamma ph+ph-1}\!\!\Psi(t)\Big]_{s_{z}s_1}\Lambda_{s_1}\Big[\prod_{t=\gamma ph}^{\gamma ph+h-1}\Psi(t)\Big]\\
&\leq(1-\epsilon^{(p-1)h})+\epsilon^{(p-1)h}(1-\epsilon^h)\leq1-\epsilon^{ph}.
\end{aligned}
\end{equation*}
\end{small}

\vspace{-1ex}
\noindent
for $s_z=1,2,\ldots,n$. This immediately yields that
\begin{small}
\begin{equation}\label{eq:14}
\begin{aligned}
\Big\|\prod_{t=\gamma ph}^{\gamma ph+ph-1}\!\!\Psi(t)\Big\|_{\infty}\leq1-\epsilon^{ph}.
\end{aligned}
\end{equation}
\end{small}

\vspace{-1ex}
\noindent
It further follows that
\[\Big\|\prod_{t=0}^{\infty}\Psi(t)\Big\|_{\infty}\leq\prod_{\gamma=0}^{\infty}\Big\|\prod_{t=\gamma ph}^{\gamma ph+ph-1}\!\!\Psi(t)\Big\|_{\infty}=0.\]
Equivalently, Eq. (\ref{eq:9}) holds. Consequently, model (\ref{eq:8}) is written as
\begin{equation}\label{eq:15}
\begin{aligned}
\lim_{t\rightarrow\infty}x^*[t+1]=       \begin{bmatrix}
              \mathbf{0} & \lim_{t\rightarrow\infty}r^*(t) \\
              \mathbf{0} & 1 \\
            \end{bmatrix}x^*[0],
\end{aligned}
\end{equation}
where $\lim_{t\rightarrow\infty}r^*(t)\!=\!\mathbf{1}_{n-1}$. This means that $\lim_{t\rightarrow\infty}x_i[t]\!=\!-x_n[0]$, $v_i\in\mathscr{V}_1$ and $\lim_{t\rightarrow\infty}x_i[t]\!=\!x_n[0]$, $v_i\in\mathscr{V}_2$. Consequently, opinion polarization is achieved.
\end{IEEEproof}

As can be seen from Theorem \ref{theorem:1}, the self-confidence levels of all followers are assumed to be non-zero to achieve opinion polarization in the asynchronous interaction scenario. In fact, opinion polarization can still be achieved when there is no constraint on the self-confidence levels of the agents in the synchronous interaction scenario, as shown in the following corollary.

\begin{corollary}\label{corollary:1}
Let each agent update the opinion by the following synchronous model
\begin{equation*}
x_i[t+1]=\theta_ix_i[t]+(1\!-\!\theta_i)u_i[t],
\end{equation*}
where $u_i[t]$ is given in (\ref{eq:4}). Opinion polarization is achieved if the interaction network $\mathscr{G}$ is structurally balanced and contains a spanning tree rooted at the leader.
\end{corollary}

\begin{IEEEproof}
Below we consider the cases of $\theta_i\neq0$ and $\theta_i=0$, respectively. As we pointed out in Remark \ref{remark:1}, the synchronous interaction is actually a special case of the asynchronous interaction. Consequently, it it easy to realize opinion polarization when $\theta_i\neq0$ according to the proof of Theorem \ref{theorem:1}. When $\theta_i=0$, the synchronous model can be written as
\begin{equation*}
x^*[t+1]=R(t) x^*[t],
\end{equation*}
where $R(t)=[r_{ij}(t)]$ is a row-stochastic matrix with all diagonal elements being zeros, and
\begin{equation*}
\begin{aligned}
r_{ij}(t)=\frac{a_{ij}f_{ij}(t)}{\sum_{v_j\in\mathscr{N}_i(t)}f_{ij}(t)}, \ i\neq j.
\end{aligned}
\end{equation*}
In addition, $R(t)$ has the following form
\begin{equation*}
R(t)=\begin{bmatrix}
              R_1(t) & R_2(t) \\
              \mathbf{0} & 1 \\
            \end{bmatrix},
\end{equation*}
where $R_1(t)$ is a sub-stochastic matrix. Using a method similar to the analysis of sub-stochastic matrix $\Psi(t)$ in Theorem \ref{theorem:1}, we can deduce that $\prod_{t=0}^{\infty}R_1(t)=\mathbf{0}$ when $\mathscr{G}$ contains a spanning tree rooted at the leader. It thus follows that
\begin{equation*}
\prod_{t=0}^{\infty}R(t)=\begin{bmatrix}
              \mathbf{0} & \mathbf{1}_{n-1} \\
              \mathbf{0} & 1 \\
            \end{bmatrix},
\end{equation*}
which leads to further results $\lim_{t\rightarrow\infty}x_i[t]=-x_n[0]$, $i=1,2,\ldots,m$ and $\lim_{t\rightarrow\infty}x_i[t]=x_n[0]$, $i=m+1,m+2,\ldots,n$. Equivalently, opinion polarization is achieved.
\end{IEEEproof}

\section{Discussion on opinion consensus}\label{section:5}

Consensus is another basic evolutionary result of different opinions held by social agents on certain specific events besides polarization. To date, related studies \cite{Altafini2013Consensus,Xia2016Structural,Proskurnikov2016Opinion,Liu2019Polar} on the Altafini model have obtained the results of opinion consensus on the signed social network without opinion leaders. However, in many real-world social networks where there exists an opinion leader, the powerful influence of the opinion leader will drive all agents to achieve opinion consensus even if trust and distrust coexist in the interactions among agents. For example, in the film ``12 Angry Men", juror 8 who believes that the boy is innocent finally persuades the rest of the jurors  to agree with him by presenting sufficient evidences again and again. Inspired by this, this section aims to analyze the consensus dynamics of signed social networks with asynchronous evolution of trust/distrust level based on opinion difference, in the presence of an opinion leader.


An indisputable fact is that when using the distributed interaction rule (\ref{eq:4}), the agents' opinions cannot gradually reach consensus under the setting of asynchronous evolution of trust/distrust level, which is theoretically confirmed by the following theorem.

\begin{theorem}\label{theorem:2}
Consider model (\ref{eq:3}) with the interaction rule (\ref{eq:4}). If the signed social network $\mathscr{G}$ contains a spanning tree rooted at the leader and $\theta_i\neq0$, $i=1,2,\ldots,n-1$, then the opinions of the followers gradually converge to the convex combinations of the opinion and opposite opinion of the leader, i.e.,
\begin{eqnarray}\label{eq:16}
\begin{aligned}
\lim_{t\rightarrow\infty}x_{i}[t]&=c_{i1}x_n[0]+c_{i2}(-x_n[0]), \\
i&=1,2,\ldots,n-1,
\end{aligned}
\end{eqnarray}
where $c_{i1}\ge 0, c_{i2}\ge 0$, and $c_{i1}+c_{i2}=1$. Furthermore, there exists at least one $i\in\{1,2,\ldots,n-1\}$ such that $c_{i2}\neq0$, that is, opinion consensus is not achieved unless $x_n[0]=0$.
\end{theorem}

\begin{IEEEproof}
Using the interaction rule (\ref{eq:4}), model (\ref{eq:3}) can first be converted equivalently to a vector-space form
\begin{eqnarray}\label{eq:17}
Y[t+1]=\Upsilon(t)Y[t],
\end{eqnarray}
where
\begin{equation*}
\begin{aligned}
&Y[t]=\big[x_{1}[t],-x_{1}[t],\cdots,x_{n}[t],-x_{n}[t]\big]^T, \\
&\Upsilon(t)=\tilde{\Theta}(t)+\big(I_{2n}-\tilde{\Theta}(t)\big)N(t),
\end{aligned}
\end{equation*}
in which $\tilde{\Theta}(t)\!=\!{\rm diag}\big\{\theta_1(t),\theta_1(t),\!\ldots\!,\theta_{n}(t),\theta_{n}(t)\big\}$ and the elements of $N(t)$ satisfy: if $q_{ij}(t)\geq0$, then
\begin{equation*}
\begin{aligned}
&\big[N(t)\big]_{2i-1,2j-1}=\big[N(t)\big]_{2i,2j}=q_{ij}(t), \\
&\big[N(t)\big]_{2i-1,2j}=\big[N(t)\big]_{2i,2j-1}=0,
\end{aligned}
\end{equation*}
and if $q_{ij}(t)<0$, then
\begin{equation*}
\begin{aligned}
&\big[N(t)\big]_{2i-1,2j}=\big[N(t)\big]_{2i,2j-1}=-q_{ij}(t), \\
&\big[N(t)\big]_{2i-1,2j-1}=\big[N(t)\big]_{2i,2j}=0.
\end{aligned}
\end{equation*}
Under the condition $\theta_i\neq 0$, $i=1,2,\ldots,n-1$, $\Upsilon(t)$ is a row-stochastic matrix with positive diagonal elements. In addition, $\Upsilon(t)$ has the following block-matrix form
\begin{equation*}
\Upsilon(t)=\left[
              \begin{array}{cc}
                \Upsilon_{1}(t)& m(t) \\
                \mathbf{0} & I_2 \\
              \end{array}
            \right],
\end{equation*}
where $\Upsilon_1(t)$ is a sub-stochastic matrix with positive diagonal elements. Using the analysis method similar to the sub-stochastic matrix $\Psi(t)$ in Theorem \ref{theorem:1}, we can derive that $\prod_{t=0}^{\infty}\Upsilon_1(t)=\mathbf{0}$. It follows that
\begin{eqnarray*}
\lim_{t\rightarrow\infty}Y[t]=\left[
              \begin{array}{cc}
                \mathbf{0}& \lim_{t\rightarrow\infty}m^*(t) \\
                \mathbf{0} & I_2 \\
              \end{array}
            \right]Y[0],
\end{eqnarray*}
where $m^*(t)\!=\!\sum_{i=0}^{t}\Upsilon_1(t)\cdots \Upsilon_1(i\!+\!1)m(i)$ is a row-stochastic matrix. Equivalently, we have $\lim_{t\rightarrow\infty}x_{i}[t]\!=\!c_{i1}x_n[0]\!+\!c_{i2}(\!-\!x_n[0])$ for $i\!=\!1,2,\ldots,n\!-\!1$, where $c_{i1}\!+\!c_{i2}\!=\!1$ and
\[c_{i1}=\lim_{t\rightarrow\infty}[m^*(t)]_{2i-1,1}=\lim_{t\rightarrow\infty}[m^*(t)]_{2i,2},\] \[c_{i2}=\lim_{t\rightarrow\infty}[m^*(t)]_{2i-1,2}=\lim_{t\rightarrow\infty}[m^*(t)]_{2i,1}.\]

Below we prove that opinion consensus is not achieved. Since $\mathscr{G}$ has a spanning tree rooted at the leader, we can find a follower $v_{s_{z}}$ such that there exists a directed path $\mathscr{P}_{v_n\rightarrow v_{s_z}}\!=\!v_n\!\rightarrow \!v_{s_{1}}\!\rightarrow\!\cdots\!\rightarrow \!v_{s_{z}}$ with both positive and negative edges. Without loss of generality, we assume that the edges $(v_n,v_{s_1}),(v_{s_1},v_{s_2}),\ldots,(v_{s_{r-1}},v_{s_{r}})$ have negative signs and the edges $(v_{s_{r}},v_{s_{r+1}}),\ldots,(v_{s_{z-1}},v_{s_{z}})$ have positive signs, where $r<z$. When $s_{r}$ is an odd and an even, respectively, the following paths can be found in $\mathscr{G}(t)$,
\begin{eqnarray*}
\begin{aligned}
&\mathscr{P}_{2n\rightarrow 2s_z-1}=2n\rightarrow 2s_1-1\rightarrow 2s_2\rightarrow\cdots\rightarrow2s_r-1\\
& \ \ \ \ \ \ \ \ \ \ \ \ \ \ \ \ \ \ \rightarrow2s_{r+1}-1\rightarrow\cdots2s_z-1,\\
&\mathscr{P}_{2n-1\rightarrow 2s_z}=2n-1\rightarrow 2s_1\rightarrow 2s_2-1\rightarrow\cdots\rightarrow2s_r\\
& \ \ \ \ \ \ \ \ \ \ \ \ \ \ \ \ \ \ \rightarrow2s_{r+1}\rightarrow\cdots2s_z.
\end{aligned}
\end{eqnarray*}
This, along with $[N(t)]_{2s_1,2n-1}>0$ and $[N(t)]_{2s_1-1,2n}>0$ which are obtained by $q_{ns_1}<0$, guarantees that $c_{s_z2}=\lim_{t\rightarrow\infty}[m^*(t)]_{2s_z-1,2}=\lim_{t\rightarrow\infty}[m^*(t)]_{2s_z,1}>0$. This means that the final opinion of follower $v_{s_z}$ cannot agree with the leader's opinion, namely, opinion consensus is not achieved unless $x_n[0]=0$.
\end{IEEEproof}

As we can see in Theorem \ref{theorem:2}, the Altafini-style interaction rule (\ref{eq:4}) cannot lead to opinion consensus on the signed social network when the leader holds a non-zero initial opinion. To analyze the evolution mechanism of opinion consensus  when distrust relations exist in the social network, such as the scenario in the film ``12 Angry Men", we next turn to a nonlinear DeGroot-style distributed interaction rule:
\begin{equation}\label{eq:18}
u_i[t]=\sum_{v_j\in\mathscr{N}_i}\frac{a_{ij}f_{ij}(t)}{\sum_{v_j\in\mathscr{N}_i}a_{ij}f_{ij}(t)}x_j[t].
\end{equation}
Based on the construction of digraph $\mathscr{G}(t)$, we can rearrange model (\ref{eq:3}) as
\begin{equation}\label{eq:19}
\begin{aligned}
x_i[t\!+\!1]=\theta_i(t)x_i[t]\!+\!\big(1\!-\!\theta_i(t)\big)\sum_{v_j\in\mathscr{N}_i(t)}p_{ij}(t),
\end{aligned}
\end{equation}
where
\begin{equation}\label{eq:20}
\begin{aligned}
p_{ij}(t)=\frac{a_{ij}(t)f_{ij}(t)}{\sum_{v_j\in\mathscr{N}_i(t)}a_{ij}(t)f_{ij}(t)}.
\end{aligned}
\end{equation}

Let $x[t]=\big[x_i[t],x_2[t],\ldots,x_n[t]\big]^T$, then (\ref{eq:19}) can be expressed as the following state-space form
\begin{equation}\label{eq:21}
x[t+1]=\Gamma(t)x[t],
\end{equation}
where $\Gamma(t)=\Theta(t)+\big(I_n-\Theta(t)\big)P(t)$ with $P(t)\!=\![p_{ij}(t)]$ and $\Theta(t)\!=\!{\rm diag}\big\{\!\theta_1(t),\!\ldots\!,\theta_{n-1}(t),1\!\big\}$. Observing the construction of $\Gamma(t)$, it is known that $\Gamma(t)$ is a general row-stochastic matrix. And hence, it is necessary to analyze the infinite products of general row-stochastic matrices $\Gamma(t)$, $t\in\mathbb{N}$ in order to investigate the conditions for (\ref{eq:21}) to achieve opinion consensus.

Before proceeding, the following notations need to be introduced,
\begin{equation*}
\begin{aligned}
&|\mathscr{N}^-_{max}|=\max\{|\mathscr{N}^-_{i}|\mid i=1,2,\ldots,n-1\},\\
&|\mathscr{N}^+_{min}|=\min\{|\mathscr{N}^+_{i}|\mid i=1,2,\ldots,n-1\},\\
&|\mathscr{N}^-_{min}|=\min\{|\mathscr{N}^-_{i}|\mid i=1,2,\ldots,n-1\},\\
&|\mathscr{N}^+_{max}|=\max\{|\mathscr{N}^+_{i}|\mid i=1,2,\ldots,n-1\}.
\end{aligned}
\end{equation*}
Now, we show in the following theorem that the DeGroot-style interaction rule (\ref{eq:18}) can effectively guarantee the realization of opinion consensus on the signed social network in presence of an opinion leader.

\begin{theorem}\label{theorem:3}
Consider model (\ref{eq:3}) with the interaction rule (\ref{eq:18}). Opinion consensus can be achieved if the interaction network  $\mathscr{G}$ contains a spanning tree $\mathscr{T}$ with the root being the leader and all edges representing trust relationships, and the following condition holds
\begin{equation}\label{eq:22}
\begin{aligned}
l^{ph}-\sigma^{ph}\!<\!1,
\end{aligned}
\end{equation}
where $p$ is the longest distance from the leader to the followers on the spanning tree $\mathscr{T}$ and
\begin{small}
\begin{equation*}
\begin{aligned}
&l\!=\!1\!+\!\frac{2(1\!-\!\theta_{min})|\mathscr{N}^-_{max}|\max\{\overline{\iota},\overline{\kappa}\}}
{|\mathscr{N}^+_{min}|\min\{\underline{\alpha},\underline{\beta}\}-|\mathscr{N}^-_{max}|\max\{\overline{\iota},\overline{\kappa}\}},\\
&\sigma\!=\!\min\!\left\{\!\theta_{min},\frac{(1\!-\!\theta_{max})\min\{\underline{\alpha},\underline{\beta}\}}
{|\mathscr{N}^+_{max}|\min\{\underline{\alpha},\underline{\beta}\}\!-\!|\mathscr{N}^-_{min}|\max\{\overline{\iota},\overline{\kappa}\}}\!\right\},
\end{aligned}
\end{equation*}
\end{small}

\vspace{-1ex}
\noindent
with $|\mathscr{N}^+_{min}|\min\{\underline{\alpha},\underline{\beta}\}>|\mathscr{N}^-_{max}|\max\{\overline{\iota},\overline{\kappa}\}$.
\end{theorem}

\begin{IEEEproof}
According to the division of the opinion leader and the followers, $\Gamma(t)$ has a block-matrix form as follows
\begin{equation*}
\Gamma[t]=\begin{bmatrix}
              \Phi(t) & w(t) \\
              \mathbf{0} & 1 \\
            \end{bmatrix},
\end{equation*}
in which
\begin{equation*}
\begin{aligned}
&\Phi(t)\!=\!\Theta^*(t)+\big(I_{n-1}\!-\!\Theta^*(t)\big)P(t)(1:n\!-\!1,1:n\!-\!1),\\
&w(t)\!=\!\big(I_{n-1}\!-\!\Theta^*(t)\big)P(t)(1:n\!-\!1,n:n).
\end{aligned}
\end{equation*}
Since $\Gamma(t)$ is a general row-stochastic matrix, it is known from Lemma \ref{lemma:2} that $\prod_{t=0}^{\infty}\Gamma(t)$ is still a general row-stochastic matrix. Thus, if the following equation holds
\begin{equation}\label{eq:23}
\begin{aligned}
\prod_{t=0}^{\infty}\Phi(t)=\mathbf{0},
\end{aligned}
\end{equation}
we have $\lim_{t\rightarrow\infty}w^*(t)\!=\!\mathbf{1}_{n-1}$, which implies the realization of opinion consensus. Therefore, our purpose below is to prove Eq. (\ref{eq:23}). It should be noted that $\Phi(t)$ is a matrix containing both positive and negative elements. According to the known property of matrix infinite norm, that is, $\|AB\|_{\infty}\!\leq\!\||A||B|\|_{\infty}$, we have
\begin{equation*}
\begin{aligned}
\Big\|\prod_{t=0}^{\infty}\Phi(t)\Big\|_{\infty}\leq\Big\|\prod_{t=0}^{\infty}|\Phi(t)|\Big\|_{\infty},
\end{aligned}
\end{equation*}
and thus, it is sufficient to prove
\begin{equation}\label{eq:24}
\begin{aligned}
\Big\|\prod_{t=0}^{\infty}|\Phi(t)|\Big\|_{\infty}=0
\end{aligned}
\end{equation}
for ensuring that Eq. (\ref{eq:23}) holds. By (\ref{eq:22}), it can be calculated that $\theta_{min}>0$, which guarantees that $|\Phi(t)|$ is a nonnegative matrix with the positive diagonal elements satisfying $[|\Phi(t)|]_{ii}>\sigma$, $i=1,2,\ldots,n-1$. Furthermore, the non-diagonal elements of $|\Phi(t)|$ satisfy:
\begin{equation*}
[|\Phi(t)|]_{ij}\!=\!\left\{\!\!
\begin{aligned}
&g^*\Big(a_{ij}(t),\theta_{i}(t),f_{ij}(t)\Big)>\sigma, \ if \ a_{ij}(t)\neq0,\\
&g^*\Big(a_{ij}(t),\theta_{i}(t),f_{ij}(t)\Big)=0, \ if \ a_{ij}(t)=0.
\end{aligned}
\right.
\end{equation*}
Through calculation, the row sums of $|\Phi(t)|$ satisfy: if $t\in\{s_k^i\}$ and $(v_n,v_i)\in \mathscr{E}(\mathscr{T})$, that is $a_{ij}(t)\neq0$, then
\begin{equation*}
\begin{aligned}
\Lambda_i[|\Phi(t)|]&\leq l-g^*\Big(a_{in}(t),\theta_{i}(t),f_{in}(t)\Big)\\
&\leq l-\sigma<1,
\end{aligned}
\end{equation*}
and if $t\in\{s_k^i\}$ and $(v_n,v_i)\notin \mathscr{E}(\mathscr{T})$, or $t\notin\{s_k^i\}$, namely, $a_{ij}(t)=0$, then \[1\leq\Lambda_i[|\Phi(t)|]\leq l,\]
where $\mathscr{E}(\mathscr{T})$ denotes the edge set of the spanning tree $\mathscr{T}$. Consequently, $|\Phi(t)|$ is a \emph{super-stochastic} matrix.

In order to get Eq. (\ref{eq:24}), we first make an isometric division of the time axis, i.e., $[0,ph),[ph,2ph),[2ph,3ph),\ldots$. Then, we prove $\|\prod_{t=\gamma ph}^{\gamma ph+ph-1}\Phi(t)\|_{\infty}<1$ associated with each time interval $[\gamma ph,\gamma ph+ph-1)$, where $\gamma\in\mathbb{N}$. Since $\mathscr{G}$ contains a spanning tree rooted at the leader, there exists a directed path $\mathscr{P}_{v_n\rightarrow v_{s_z}}\!=\!v_n\!\rightarrow \!v_{s_{1}}\!\rightarrow\!\cdots\!\rightarrow \!v_{s_{z}}$ for each follower $v_{s_{z}}$. According to (\ref{eq:1}), one knows that each agent updates its opinion at least once in arbitrary time interval with length $h$. Thus, in the interval $[\gamma ph,\gamma ph+h)$, there exists a time $\gamma ph+l_1$ such that $(v_n,v_{s_{1}})\in\mathscr{E}(\gamma ph+l_1)$, which leads to
\begin{equation}\label{eq:25}
\begin{aligned}
\Lambda_{s_1}\big[ |\Phi(\gamma ph+l_1)|\big]\leq l-\sigma.
\end{aligned}
\end{equation}
By the result in Lemma \ref{lemma:1}, we have
\begin{small}
\begin{equation}\label{eq:26}
\begin{aligned}
\Lambda_{j}\Big[\prod_{t=\gamma ph}^{\gamma ph+l_1-1}|\Phi(t)|\Big]\leq l^{l_1}, \ j=1,2,\ldots,n-1.
\end{aligned}
\end{equation}
\end{small}

\vspace{-1ex}
\noindent
Combining (\ref{eq:25}) and (\ref{eq:26}), it can be derived that
\begin{small}
\begin{equation}\label{eq:27}
\begin{aligned}
\Lambda_{s_1}\Big[\prod_{t=\gamma ph}^{\gamma ph+l_1}\!\!|\Phi(t)|\Big]&\!=\!\sum_{j}\big[|\Phi(\gamma ph\!+\!l_1)|\big]_{s_1j}\Lambda_{j}\Big[\!\prod_{t=\gamma ph}^{\gamma ph+l_1-1}\!\!|\Phi(t)|\Big]\\
&\leq \sum_{j}\big[|\Phi(\gamma ph\!+\!l_1)|\big]_{s_1j}l^{l_1}\leq (l-\sigma)l^{l_1}.
\end{aligned}
\end{equation}
\end{small}

\vspace{-1ex}
\noindent
Noting the fact $[|\Phi(t)|]_{s_1s_1}\geq\sigma$ for any $t\in\mathbb{N}$, there holds
\begin{small}
\begin{equation}\label{eq:28}
\begin{aligned}
\Big[\prod_{t=\gamma ph+l_1+1}^{\gamma ph+h-1}\!\!|\Phi(t)|\Big]_{s_1s_1}\geq \sigma^{h-l_1-1}.
\end{aligned}
\end{equation}
\end{small}

\vspace{-1ex}
\noindent
According to (\ref{eq:27}) and (\ref{eq:28}), the following result is obtained
\begin{small}
\begin{equation}\label{eq:29}
\begin{aligned}
\Lambda_{s_1}\!\Big[\!\prod_{t=\gamma ph}^{\gamma ph+h-1}\!\!|\Phi(t)|\Big]&\!=\!\sum_{j}\!\Big[\prod_{t=\gamma ph+l_1+1}^{\gamma ph+h-1}\!\!|\Phi(t)|\Big]_{s_1j}\!\Lambda_{j}\!\Big[\prod_{t=\gamma ph}^{\gamma ph+l_1}\!\!|\Phi(t)|\Big]\\
& \ \ +\!\Big[\prod_{t=\gamma ph+l_1+1}^{\gamma ph+h-1}\!\!|\Phi(t)|\Big]_{s_1s_1}\!\!\Lambda_{s_1}\!\Big[\prod_{t=\gamma ph}^{\gamma ph+l_1}\!\!|\Phi(t)|\Big]\\
&\!\leq\!(l^{h-l_1\!-\!1}\!\!-\!\sigma^{h-l_1\!-\!1})l^{l_1\!+\!1}\!+\!\sigma^{h-l_1-1}(l\!-\!\sigma)l^{l_1}\\
&\!\leq\!l^h-\sigma^{h-l_1}l^{l_1}\!\leq\!l^h-\sigma^h.
\end{aligned}
\end{equation}
\end{small}

Consider the intervals $[\gamma ph+yh-h,\gamma ph+yh)$, $y=2,3,\ldots,z$. In each interval, there exists a time $\gamma ph+yh-h+l_y\in[\gamma ph+yh-h,\gamma ph+yh]$ such that $(s_{y-1},s_y)\in\mathscr{E}(\gamma ph+yh-h+l_y)$, and then we have
\begin{equation}\label{eq:30}
\begin{aligned}
\big[|\Phi(\gamma ph+yh-h+l_y)|\big]_{s_ys_{y-1}}\geq\sigma.
\end{aligned}
\end{equation}
In addition, it is also known that
\begin{small}
\begin{equation}\label{eq:31}
\begin{aligned}
&\Big[\prod_{t=\gamma ph+yh-h}^{\gamma ph+yh-h+l_y-1}|\Phi(t)|\Big]_{s_{y-1}s_{y-1}}\geq\sigma^{l_y},\\
&\Big[\prod_{t=\gamma ph+yh-h+l_y+1}^{\gamma ph+yh-1}|\Phi(t)|\Big]_{s_{y}s_{y}}\geq\sigma^{h-l_y-1}.
\end{aligned}
\end{equation}
\end{small}

\vspace{-1ex}
\noindent
Combining (\ref{eq:30}) and (\ref{eq:31}), we can deduce that
\begin{small}
\begin{equation}\label{eq:32}
\begin{aligned}
\Big[\prod_{t=\gamma ph+yh-h}^{\gamma ph+yh-1}\!\!|\Phi(t)|\Big]_{s_{y}s_{y-1}}\geq\sigma^{h}.
\end{aligned}
\end{equation}
\end{small}

\vspace{-1ex}
\noindent
Equivalently,
\begin{small}
\begin{equation}\label{eq:33}
\begin{aligned}
\Big[\prod_{t=\gamma ph+h}^{\gamma ph+zh-1}|\Phi(t)|\Big]_{s_{z}s_{1}}\geq\sigma^{zh-h},
\end{aligned}
\end{equation}
\end{small}

\vspace{-1ex}
\noindent
which combines the fact $[\prod_{t=\gamma ph+zh}^{\gamma ph+ph-1}|\Phi(t)|]_{s_{z}s_{z}}\!\geq\!\sigma^{ph-zh}$  ensures that
\begin{small}
\begin{equation}\label{eq:34}
\begin{aligned}
\Big[\prod_{t=\gamma ph+h}^{\gamma ph+ph-1}|\Phi(t)|\Big]_{s_{z}s_{1}}\geq\sigma^{ph-h}.
\end{aligned}
\end{equation}
\end{small}

\vspace{-1ex}
\noindent
By (\ref{eq:29}) and (\ref{eq:34}), we get that
\begin{small}
\begin{equation*}
\begin{aligned}
\Lambda_{s_z}\!\Big[\!\prod_{t=\gamma ph}^{\gamma ph+ph-1}\!\!\!|\Phi(t)|\Big]&\!=\!\sum_{j\neq s_1}\!\!\Big[\!\prod_{t=\gamma ph+h}^{\gamma ph+ph-1}\!\!\!|\Phi(t)|\Big]_{s_{z}j}\!\Lambda_{j}\!\Big[\!\prod_{t=\gamma ph}^{\gamma ph+h-1}\!\!\!|\Phi(t)|\Big],\\
& \ \ +\!\!\Big[\!\prod_{t=\gamma ph+h}^{\gamma ph+ph-1}\!\!\!|\Phi(t)|\Big]_{s_{z}s_1}\!\Lambda_{s_1}\!\Big[\!\prod_{t=\gamma ph}^{\gamma ph+h-1}\!\!|\Phi(t)|\Big]\\
&\!\leq\!\big(l^{ph-h}-\sigma^{ph-h}\big)l^h+\sigma^{ph-h}\big(l^h-\sigma^{h}\big)\\
&\!=\!l^{ph}-\sigma^{ph}
\end{aligned}
\end{equation*}
\end{small}

\vspace{-1ex}
\noindent
for any $s_z=1,2,\ldots,n$. By condition (\ref{eq:22}), we have
\begin{small}
\begin{equation}\label{eq:35}
\begin{aligned}
\Big\|\!\prod_{t=\gamma ph}^{\gamma ph+ph-1}\!\!|\Phi(t)|\Big\|_{\infty}\leq l^{ph}-\sigma^{ph}<1.
\end{aligned}
\end{equation}
\end{small}

\vspace{-1ex}
\noindent
This immediately yields that
\begin{small}
\begin{equation*}
\begin{aligned}
\Big\|\prod_{t=0}^{\infty}|\Phi(t)|\Big\|_{\infty}\leq\prod_{\gamma=0}^{\infty}\Big\|\prod_{t=\gamma ph}^{\gamma ph+ph-1}\!\!|\Phi(t)|\Big\|_{\infty}=0.
\end{aligned}
\end{equation*}
\end{small}

\vspace{-1ex}
\noindent
Equivalently, Eq. (\ref{eq:23}) holds. Consequently, model (\ref{eq:21}) is written equivalently as
\begin{equation}\label{eq:36}
\begin{aligned}
\lim_{t\rightarrow\infty}x[t+1]&=\prod_{t=0}^\infty\begin{bmatrix}
              \Phi(t) & w^*(t) \\
              \mathbf{0} & 1 \\
            \end{bmatrix}x[0]\\
&=       \begin{bmatrix}
              \mathbf{0} & \lim_{t\rightarrow\infty}w^*(t) \\
              \mathbf{0} & 1 \\
            \end{bmatrix}x[0],
\end{aligned}
\end{equation}
where $w^*(t)\!=\!\sum_{i=0}^{t}\Phi(t)(t)\cdots\Phi(i\!+\!1)w(i)$. Since $\prod_{t=0}^{\infty}\Gamma(t)$ is a general row-stochastic matrix, we have $\lim_{t\rightarrow\infty}w^*(t)\!=\!\mathbf{1}_{n-1}$, which implies that  $\lim_{t\rightarrow\infty}x_i[t]\!=\!x_n[0]$ for $i\!=\!1,2,\ldots,n\!-\!1$. That is to say, opinion consensus is achieved.
\end{IEEEproof}

\section{Simulation results}\label{section:6}

In this section, we verify the correctness of our theoretical results by simulating the ``12 Angry Men" network and the Karate Club network.

\subsection{The ``12 Angry Men" network}\label{section:6A}

In the film ``12 Angry Men", 12 jurors were invited to decide whether a boy was guilty or not. For this example, we use the orderly increase of real numbers in the interval $[-1,1]$ to express the intensity from ``guilty" to ``no guilty", where $-1$ and $1$ means ``guilty" and ``no guilty", respectively. At the initial moment, jurors $1,3,4,7,10,12$ firmly believed that the boy was guilty, and jurors $2,5,6,9,11$ also considered the boy guilty after hesitating for a while. Accordingly, we assume that these jurors' initial opinions of whether a boy is guilty or not in this example are
\begin{equation*}
\begin{aligned}
&x_1[0]\!=\!-0.9, x_2[0]\!=\!-0.3, x_3[0]\!=\!-1, x_4[0]\!=\!-0.7, \\
&x_5[0]\!=\!-0.45, x_6[0]\!=\!-0.2, x_7[0]\!=\!-0.6, x_9[0]\!=\!-0.05,\\
&x_{10}[0]\!=\!-0.8, x_{11}[0]\!=\!-0.13, x_{12}[0]\!=\!-0.5.
\end{aligned}
\end{equation*}
In addition, juror $8$ was particularly convinced that the boy was not guilty and had sufficient evidences to support his opinion in the film. Thus, the initial opinion of juror $8$ in this example is assumed to be $x_8[0]=1$.

\begin{figure}[t]
  \centering
  \subfigure[$\mathscr{G}_1$]{
    \includegraphics[width=1.3in]{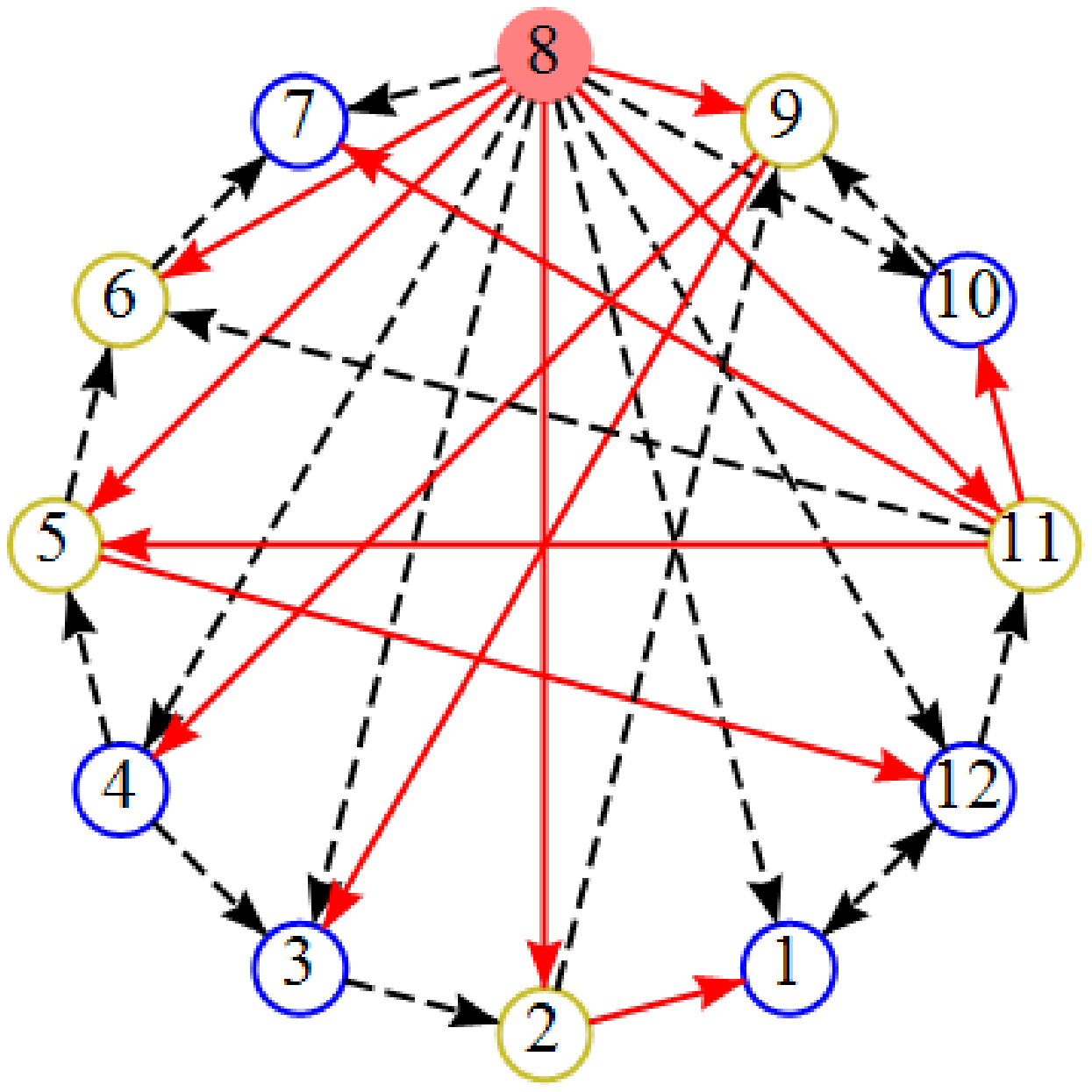}\label{fig3a}}
    \subfigure[$\mathscr{G}_2$]{
    \includegraphics[width=1.3in]{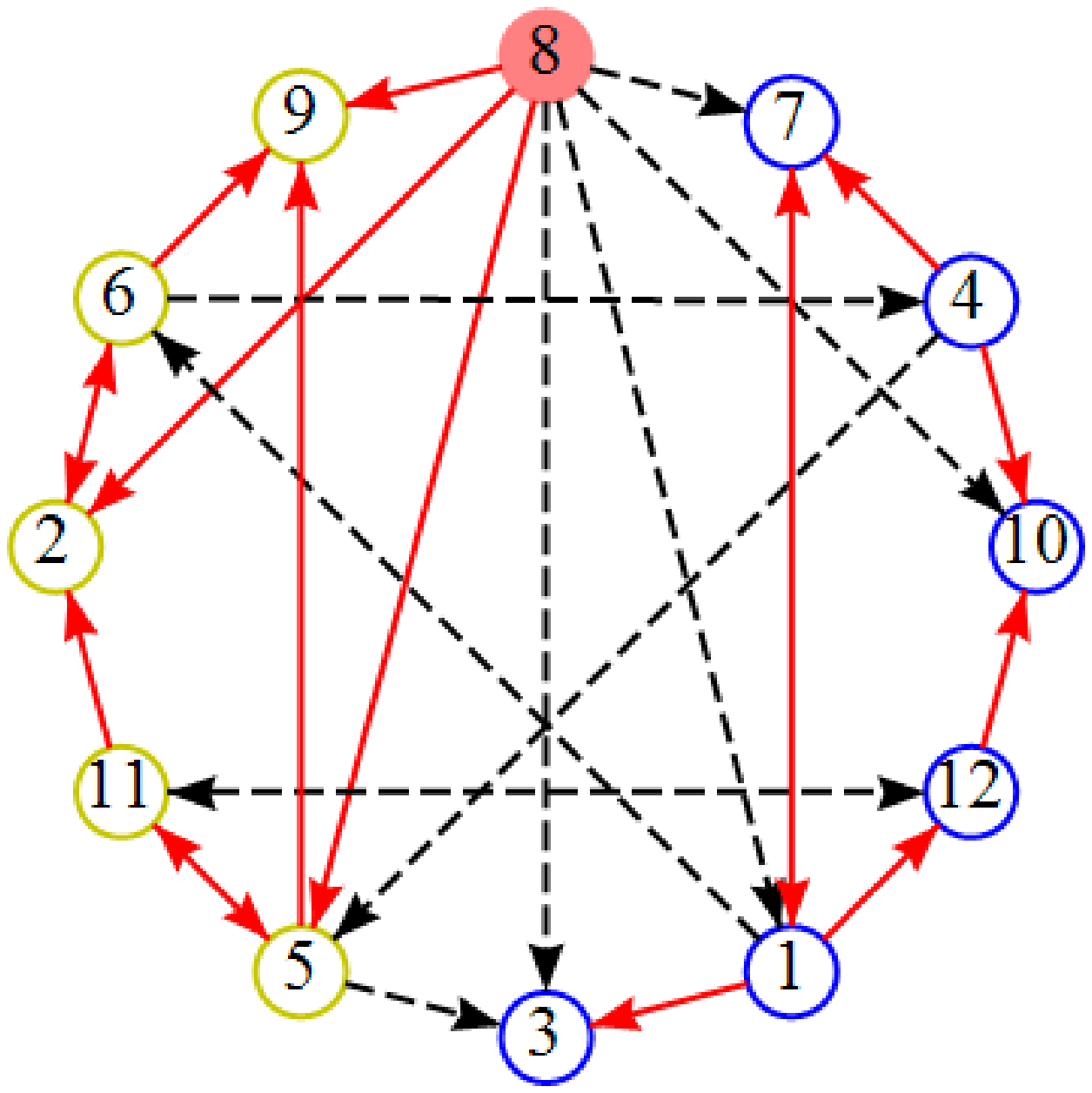}\label{fig3b}}
  \caption{Two different ``12 Angry Men" networks. The trust relations among the jurors are represented by the solid edges and the distrust relations are represented by the dotted edges. The level of trust or distrust is indicated by the weights of different color edges. }\label{fig3}
\end{figure}

According to the movie plot, we construct a directed network $\mathscr{G}_1$ shown in Fig. \ref{fig3a} to reflect as succinctly as possible the trust and distrust relationships during the discussions. A directed solid edge $(i,j)$ which has a positive sign means that juror $j$ trusts juror $i$. A directed dashed edge $(i,j)$ with negative sign represents that juror $j$ distrusts juror $i$. Since juror 8 has sufficient evidences to support his opinion, he is not affected by other jurors, which means that there are no directed edges that end at node $8$ in $\mathscr{G}_1$. Moreover, jurors $2,5,6,9,11$ were hesitant to make decisions at the initial moment because of their insufficient evidences supporting ``guilty", which indirectly indicated that they were more likely to agree with the opinion of jury $8$. The fact implies that these edges $(8,2),(8,5),(8,6),(8,9),(8,11)$ have positive signs. Furthermore, jurors $1,3,4,7,10,12$ may not be convinced of the opinion of jury $8$ since they were more determined when they initially made the ``guilty" decision. To describe this situation, we assume that the directed edges $(8,1),(8,3),(8,4),(8,7),(8,10),(8,12)$ have negative signs.

\begin{figure}[t]
  \centering
  \subfigure[distrust level of juror 3 to 8]{
    \includegraphics[width=1.65in]{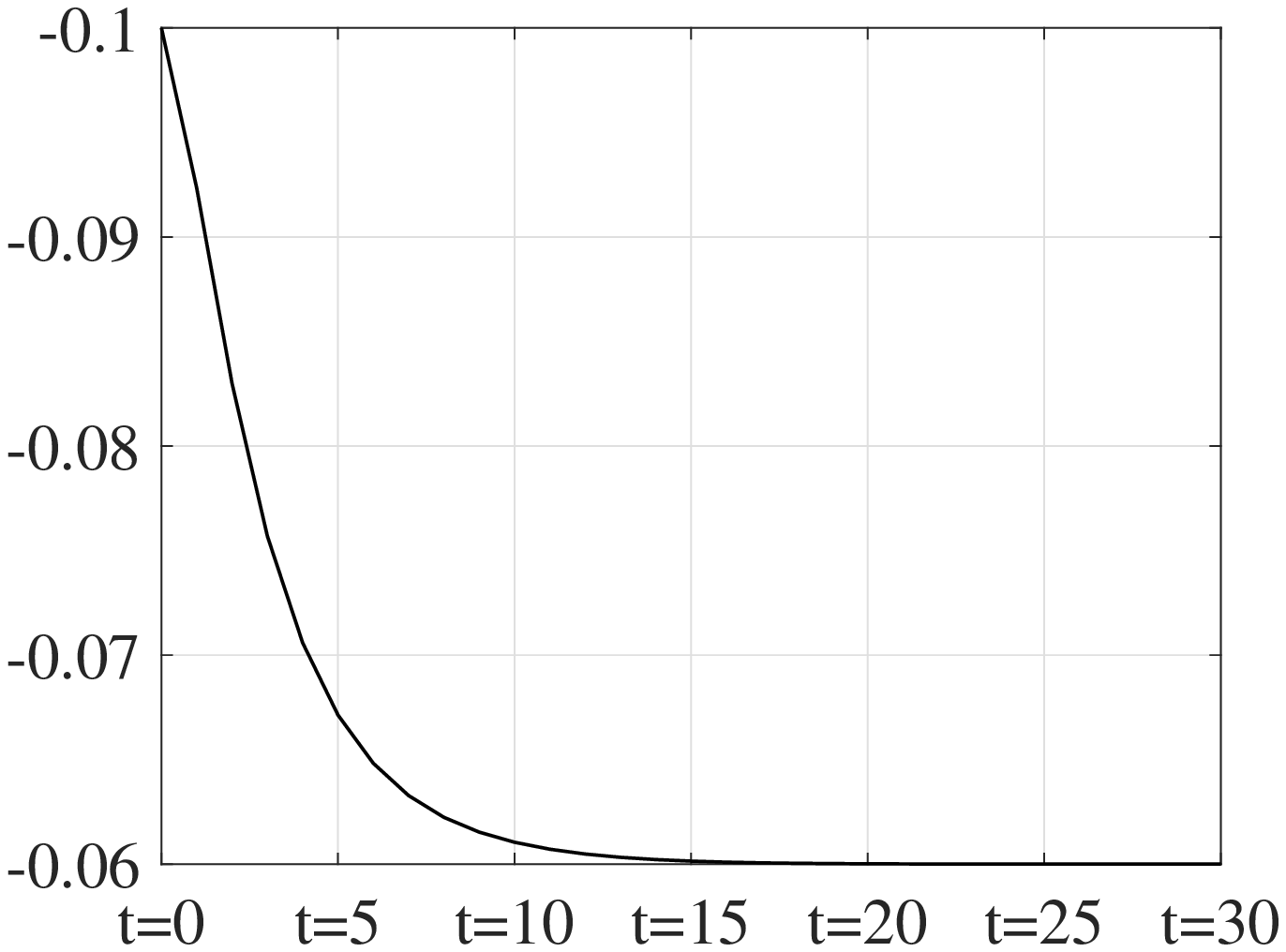}\label{fig4a}}
  \subfigure[trust level of juror 9 to 8]{
    \includegraphics[width=1.65in]{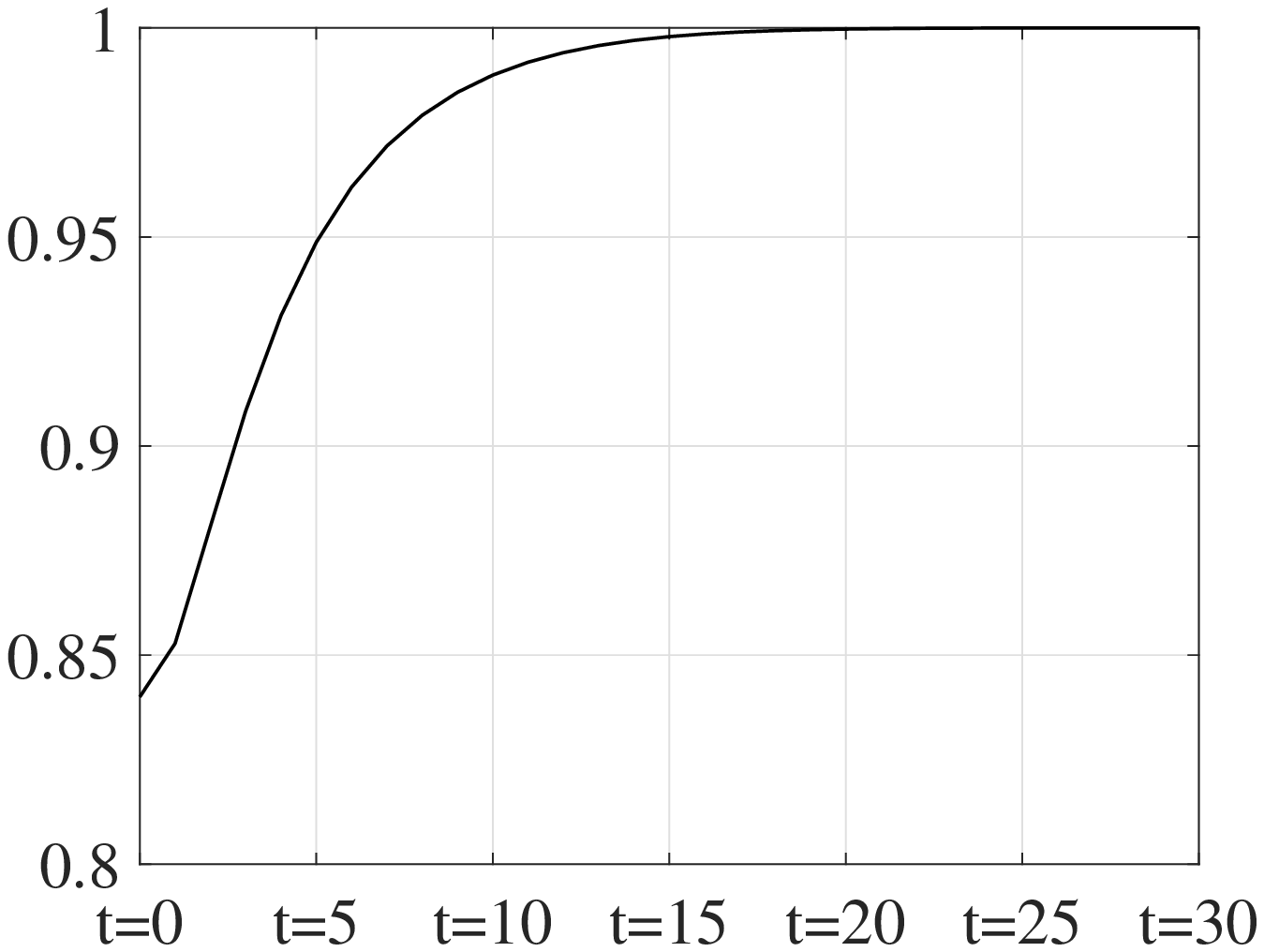}\label{fig4b}}
  \caption{Evolution of the trust/distrust levels of jurors 3 and 9 to juror 8.}\label{fig4}
\end{figure}

\begin{figure}[t]
  \centering
  \subfigure[]{
    \includegraphics[width=7.5cm,height=2.7cm]{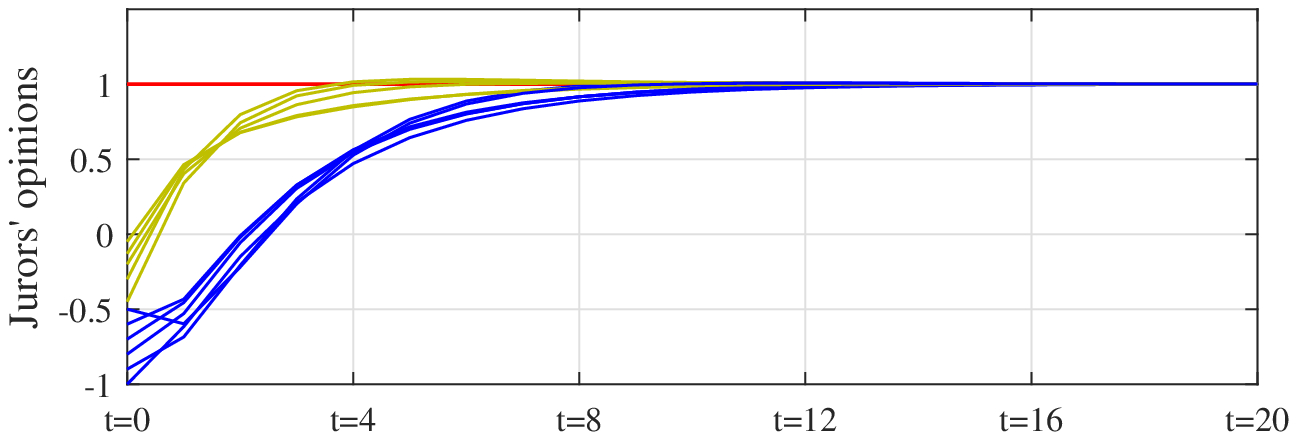}\label{fig5a}}
    \subfigure[]{
    \includegraphics[width=7.5cm,height=3cm]{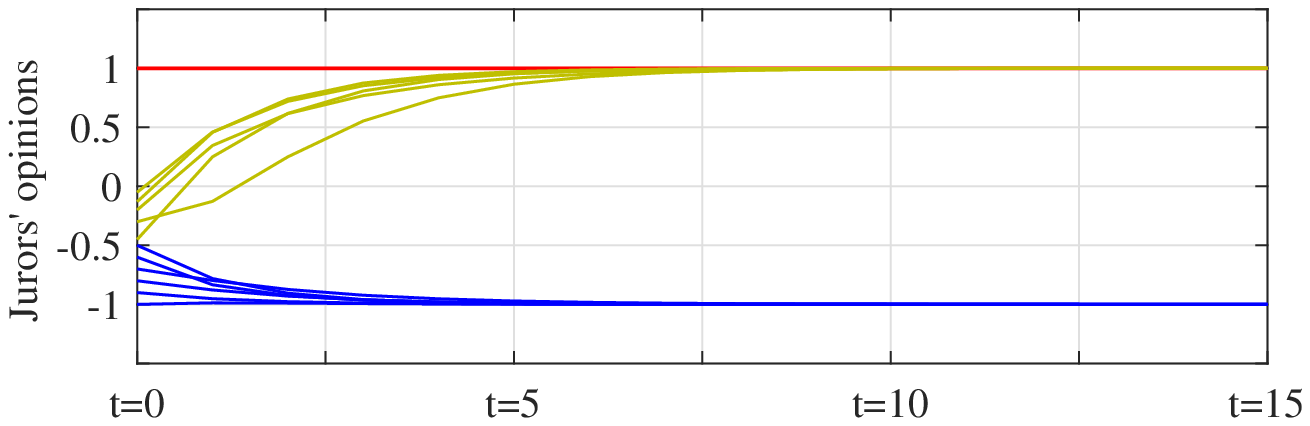}\label{fig5b}}
  \caption{Opinions evolution trajectories of the jurors interacting by different networks $\mathscr{G}_1$ and $\mathscr{G}_2$, in which red line -- trajectory of juror 8; yellow lines -- trajectories of jurors $2,5,6,9,11$; blue lines -- trajectories of jurors $1,3,4,7,10,12$. }\label{fig5}
\end{figure}

Each juror adjusts his opinion by adopting dynamic model (\ref{eq:3}) with the distributed interaction rule (\ref{eq:18}). Because each juror actively communicates with other jurors about whether the boy is guilty, we assume that each juror can be influenced by the opinions of his neighbors at any time, that is, the parameter $h$ is set as 1 in the asynchronous interaction. Suppose that the self-confidence level of juror 8 is $\theta_8=1$, while the self-confidence levels of the remaining jurors are set to $\theta_i=0.5$, $i\neq8$. In $\mathscr{G}_1$, red solid lines and black dotted lines represent trust and distrust relationships, respectively. In addition, the weights of red solid lines are quantified by a positive, decreasing, bounded function
\[f^+_{ij}(|x_i[t]\!-\!x_j[t]|)\!=\!1-0.1|x_i[t]\!-\!x_j[t]|,\]
and the weights of black dotted lines are quantified by a positive, increasing, bounded function \[f^-_{ij}(|x_i[t]\!-\!x_j[t]|)\!=\!0.02|x_i[t]\!-\!x_j[t]|+0.06,\]
where $|x_i[t]-x_j[t]|$ represents the opinions difference between neighboring jurors. With the mentioned settings, the condition (\ref{eq:22}) in Theorem \ref{theorem:3} holds.

According to the plot of the film ``12 Angry Men", we can observe that juror 3 shows the strongest distrust level to juror 8 at the beginning. However, jury 3's distrust level to jury 8 gradually decreased as juror 8 stated the evidences round after round, which can be seen from Fig. \ref{fig4a}. In addition, juror 9 appeared hesitant in convicting the boy for lack of evidence at the initial moment, so he showed trust in juror 8 but with little level when juror 8 stated sufficient evidences of the boy's innocence at the outset. Along with the opinion interaction, the trust level of juror 9 to juror 8 gradually increased, which can be observed from Fig. \ref{fig4b}. Finally, all the jurors who thought the boy was guilty at the initial moment finally turned their opinions into ``no guilty". This phenomenon is shown through Fig.~\ref{fig5a}. Furthermore, we can also clearly see from Fig.~\ref{fig5a} that jurors $2,5,6,9,11$ change their opinions faster than jurors $1,3,4,7,10,12$, which is exactly the same as the details of the film.

The simulation result Fig. \ref{fig5a} under the interaction network $\mathscr{G}_1$ gives us a lot of inspiration and endless aftertaste.
Juror 8, the male protagonist of the film, strove to justify the suspect according to the clues he has mastered, by constantly reviewing and logically scrutinizing the details of other people's opinions in the opposite direction. The point is that the language of art and clever reasoning, give people the feeling of layers of peeling cocoon. Ultimately, he successfully made the whole jury impartially determine that the suspect is innocent, which is closely related to his personal ability and charisma. But if the evidence is unconvincing to support the conclusion of innocence, or if the juror's ability is insufficient to persuade the others, that happy ending in the movie may not happen. That is to say, it is an indisputable fact that there may be a phenomenon of opinion polarization for ``12 Angry Men" networks with other network structures, such as structural balance.

Suppose jurors interact with each other through a structurally balanced network $\mathscr{G}_2$. Each juror adjusts his opinion according to model (\ref{eq:3}) with the distributed interaction rule (\ref{eq:4}). In $\mathscr{G}_2$, we can observe that all jurors are divided into two subgroups $\mathscr{V}_1=\{2,5,6,8,9,11\}$ and $\mathscr{V}_2=\{1,3,4,7,10,12\}$, where there are only trust relationships within each subgroup and only distrust relationships between subgroups. The settings of edge weights and self-confidence levels of the jurors in $\mathscr{G}_2$ are the same as those in $\mathscr{G}_1$. Finally, it can be observed from Fig. \ref{fig5b} that the jurors in $\mathscr{V}_1$ gradually change their initial judgment ``guilty" to be ``no guilty" that is consistent with juror $8$, while the jurors in $\mathscr{V}_2$ still insist on the judgment of ``guilty".

\subsection{The Karate Club network}\label{section:6B}

\begin{figure}[t]
  \centering
  \subfigure[]{
    \includegraphics[width=2.3in]{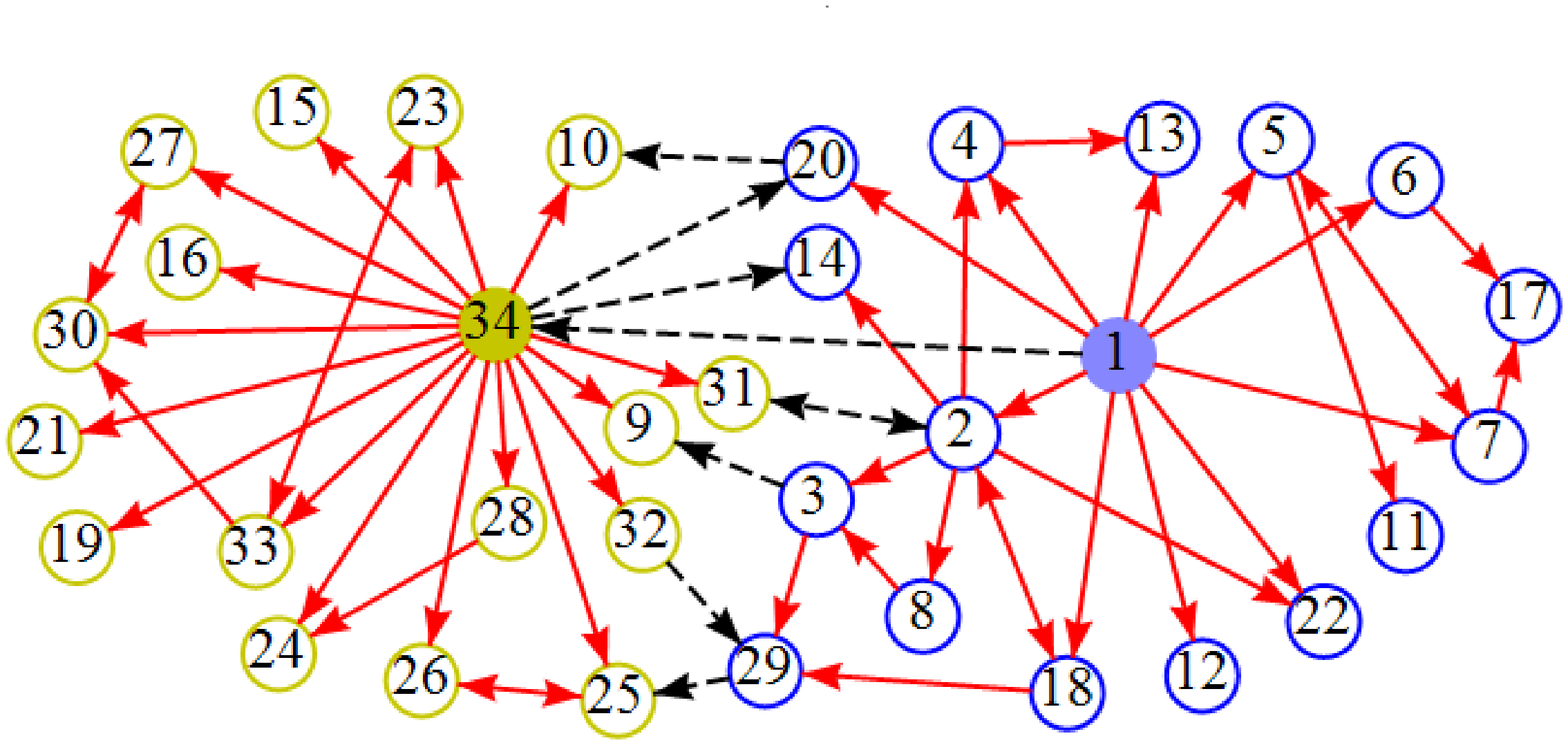}\label{fig6a}}
  \subfigure[]{
    \includegraphics[width=7.5cm,height=3cm]{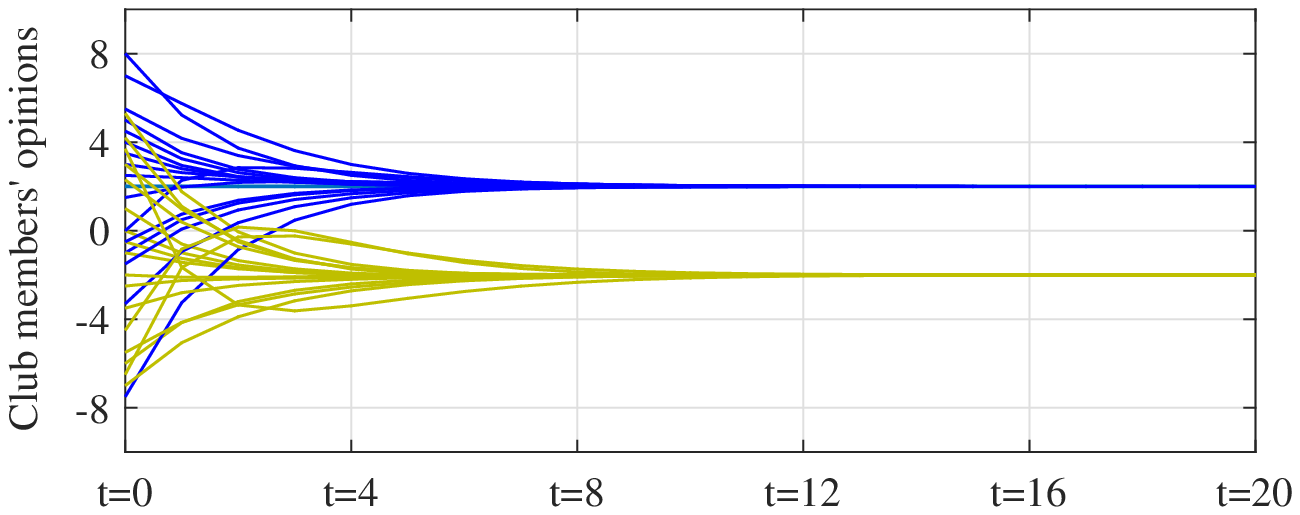}\label{fig6b}}
  \caption{The Karate Club network $\mathscr{G}_1$ and the corresponding opinion evolution trajectories of the club members. In (a), nodes 1 and 34 represent the supervisor and the coach respectively, and the solid and dotted edges describe the trust and distrust relations respectively. In (b), the yellow lines represent the opinion trajectories of the club members marked as yellow in $\mathscr{G}_1$, and the blue lines are the opinion trajectories of the club members marked as blue in $\mathscr{G}_1$. }\label{fig6}
\end{figure}

\begin{figure}[t]
  \centering
    \subfigure[]{
    \includegraphics[width=2.3in]{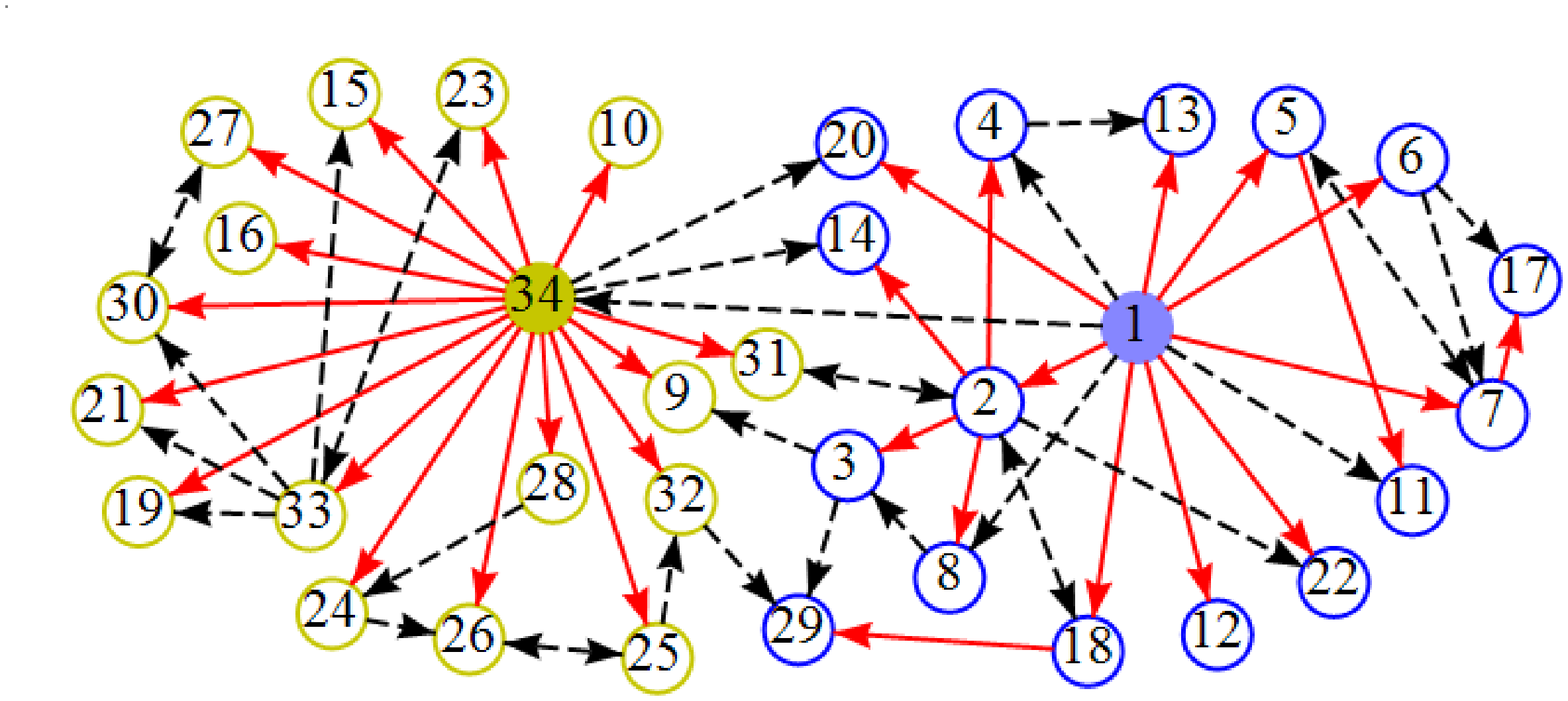}\label{fig7a}}
      \subfigure[]{
    \includegraphics[width=7.5cm,height=3cm]{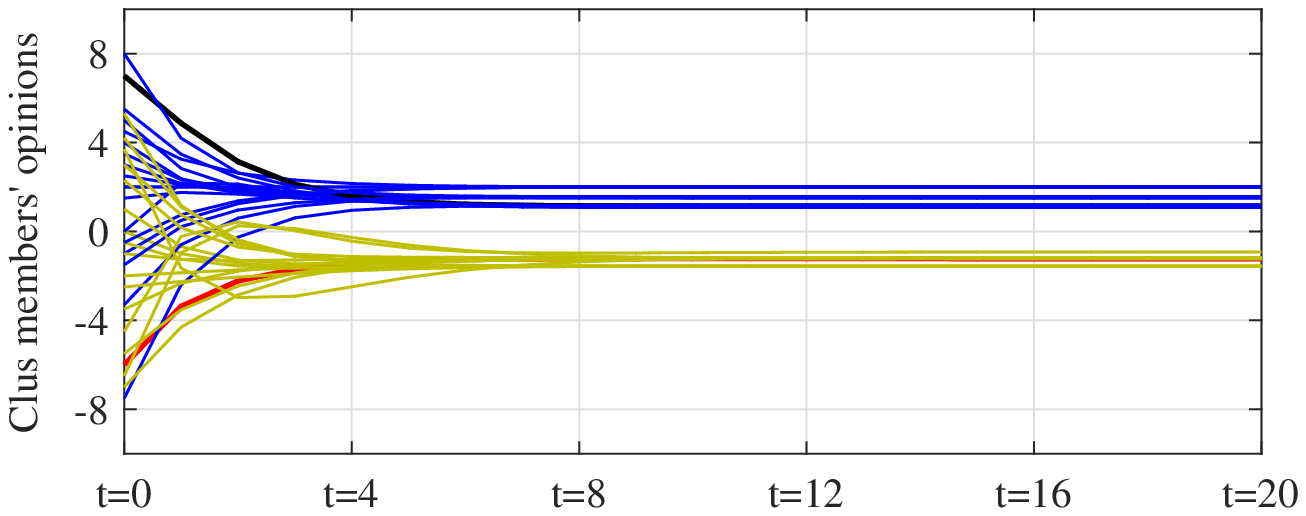}\label{fig7b}}
  \caption{The Karate Club network $\mathscr{G}_2$ and the corresponding opinion evolution trajectories of the club members. In (b), the black and red lines represent the opinion evolution trajectories of the club members 5 and 33, respectively.}\label{fig7}
 \end{figure}

The Karate Club Network proposed in \cite{Zachary1977} is a classic data set in the field of social network analysis. In the early 1970s, Zachary, a sociologist, spent two years observing the social relationships among 34 individuals of a karate club in an American university to construct a social network among the members. During this observation period, the club is split into two small groups each with its core because of the dispute between the supervisor (node 1) and the coach (node 34). Considering the polarization phenomenon in karate club, the following simulation examples are employed to verify the theoretical results of our paper based on the Karate Club network, in which each directed edge means that the member represented by the tail node has the trust relationship with the member represented by the first node.

Consider a structurally balanced Karate Club Network $\mathscr{G}_1$, shown in Fig.~\ref{fig6a}. The settings of edge weights and self-confidence levels of the club members are the same as those in the ``12 Angry Men" network. The opinion evaluation trajectories of all club members are shown in Fig. \ref{fig6b}. It is indicated from Fig. \ref{fig6b} that the club members denoted by blue circles gradually agree with the supervisor, while the opinions of others represented by purple circles finally reach the coach's opinion that is opposed to that of the supervisor.

In the actual Karate Club network, it can not be confirmed that the club members present two completely opposite subgroups in the early stage, in which the members within each subgroup completely trust each other and the members belonging to different subgroups completely distrust each other, that is, the interaction network of the club is structurally balanced. Therefore, we next use the result of Theorem \ref{theorem:2} to analyze a relatively realistic network structure, such as Fig. \ref{fig7a}, which is not structurally balanced and distrust may exist between any two members. Let $x_i[t]$ represent the opinion of the member $i$ at time $t$. The initial opinions of the supervisor and the coach are set to $x_{1}=2$ and $x_{34}=-2$, respectively, and other members' initial opinions are randomly generated in the interval $[-8,8]$, in particular, $x_5=7$ and $x_{33}=-6$. According to Eq. (\ref{eq:16}) in Theorem \ref{theorem:2}, we know that the final opinions of all members are the convex combinations of the initial opinions of the supervisor and the coach, i.e., $x_i[t]=c_{i1}x_1[0]+c_{i2}x_{34}[0]$, where $c_{i1},c_{i2}\geq0$ and $c_{i1}+c_{i2}=1$. In addition, if the convex combination coefficients of the members satisfy $c_{i1}>c_{i2}$, their opinions are closer to the supervisor's opinion (the corresponding nodes are marked as blue in Fig. \ref{fig7a}, and their opinion trajectories are shown with the blue lines in \ref{fig7b}), while their opinions are closer to that of the coach if $c_{i2}>c_{i1}$ (which are marked as yellow in \ref{fig7a} and \ref{fig7b} accordingly). For example, it can be calculated that the convex combination coefficients associated with the members $5$ and $33$ are $c_{51}=0.752$, $c_{52}=0.248$ and $c_{33,1}=0.177$, $c_{33,2}=0.823$, respectively. From $c_{51}>c_{52}$ and $c_{33,1}<c_{33,2}$, it follows that the member $5$ tends to support the supervisor, while the member $33$ trusts the coach more in this numerical example. As a result, we can also distinguish the two communities of the entire club network through the result obtained  in Theorem \ref{theorem:2}, which is described in details in Fig. \ref{fig7b}.

\section{Conclusion}\label{section:7}

In this paper, the influence of the leader on the formation of followers' opinions in signed social networks has been studied, and
an asynchronous evolution mechanism of trust/distrust level based on opinion difference has been proposed. The properties of sub-stochastic matrix and super-stochastic matrix, as well as the construction of signed digraphs, have been used to derive comprehensive theoretical results for different dynamics behaviors, e.g., opinion polarization and opinion consensus, etc. Our results complement the existing results in the literature regarding signed social networks. Moreover, the numerical simulations of opinions formation of the members in the ``12 Angry Men" network and the Karate Club network have been provided to verify the correctness of our theoretical results.

\end{document}